\documentclass{aa}  

\usepackage{graphicx}
\usepackage{xcolor}
\usepackage[export]{adjustbox}
\usepackage{pdflscape}
\usepackage{txfonts}
\usepackage[pdfpagelabels=false]{hyperref} 
\hypersetup{colorlinks=true,linkcolor=blue,citecolor=blue,filecolor=blue,urlcolor=blue,}
\usepackage{caption}
\usepackage{systeme}    
\usepackage{subcaption} 
\usepackage{breqn}      
\usepackage{rotating}
\usepackage{adjustbox}
\usepackage{tabularx}
\usepackage[flushleft]{threeparttable}

\newsavebox{\tempbox}

\newcommand{\setcaptype}[1]
  {\expandafter\def\csname @captype\endcsname{#1}\ignorespaces}

\newcommand{\dsct}{$\delta$\,Sct}
\newcommand\Tstrut{\rule{0pt}{2.6ex}}         
\newcommand{\revision}[1]{#1}
\begin{document}

   \title{Discovery of new magnetic $\delta$ Scuti stars \\ and impact of magnetism on pulsation excitation}


   \author{K. Thomson-Paressant\inst{1}\fnmsep\inst{2}\fnmsep\thanks{E-mail: keegan.thomson-paressant@obspm.fr}, 
   C. Neiner\inst{1}, 
   J. Labadie-Bartz\inst{1,3}, 
   R.-M. Ouazzani\inst{1},
   S. Mathis\inst{4} \and 
   L. Manchon\inst{1}
   }

   \institute{$^1$LIRA, Observatoire de Paris, Universit\'e PSL, CNRS, Sorbonne Universit\'e, Universit\'e Paris Cit\'e, CY Cergy Paris Universit\'e, 92190 Meudon, France \\
   $^2$School of Mathematics, Statistics and Physics, Newcastle University, Newcastle upon Tyne, NE1 7RU, UK \\
   $^3$DTU Space, Technical University of Denmark, Elektrovej 327, Kgs., Lyngby 2800, Denmark \\
   $^4$Université Paris-Saclay, Université Paris Cité, CEA, CNRS, AIM, F-91191 Gif-sur-Yvette, France}
   
   \date{Received XXX; accepted XXX}

 
  \abstract
   {At this time, the list of known magnetic $\delta$ Scuti stars is extremely limited, with only a handful of well-studied examples.}
   {We seek to expand this list, by retrieving targets from a variety of sources and demonstrating that they present simultaneously a surface magnetic field signature and $\delta$ Scuti pulsations.}
   {We obtained archival and new spectropolarimetric datasets for a variety of known $\delta$ Scuti stars and analysed them using the Least Squares Deconvolution method to generate mean Stokes I and V profiles for each target, from which we can determine longitudinal magnetic field measurements. Additionally, we assessed photometric data from the TESS satellite to discern frequency peaks consistent with $\delta$ Scuti pulsations in known magnetic stars, and to identify magnetic candidates via rotational modulation.}
   {We present a compiled list of all the confirmed magnetic $\delta$ Scuti stars discovered to date, containing 13 stars. The majority of this sample lies outside the usual $\delta$ Scuti instability strip in the H-R diagram,  though we do not observe any specific correlations between magnetic field strength and various stellar parameters. This indicates that strong global magnetic fields play a fundamental role in shaping interior structure and processes. Magnetic fields thus must be included in realistic stellar models in order to more accurately predict structure and evolution.}
   {This work constitutes the largest database to date of strongly magnetic $\delta$ Scuti stars, one that will continue to grow over time with subsequent studies. }


    \titlerunning{State-of-the-art in \ensuremath{\delta} Scuti magnetism}
    \authorrunning{K. Thomson-Paressant et al.}
   \maketitle
%

\section{Introduction}

The field of stellar magnetism has progressed in leaps and bounds in the past decade, as our techniques for identifying \emph{a priori} indicators of the presence of magnetic fields have evolved and improved over time, thanks in part to the ever-growing list of known and well-studied magnetic stars. At the same time, the field of asteroseismology has made enormous strides in quantifying the internal structure and properties of stars over a wide range of masses and ages. Stars with directly observed magnetic fields which are also pulsators are thus ideal test-beds for advancing our knowledge of the roles of magnetism in the lives of stars. 

In the case of $\delta$\,Scuti (\dsct{}) stars, the first spectropolarimetrically confirmed magnetic star was only discovered in 2013 \citep[HD\,188774, ][]{lampens2013}, with an additional few being added to the list in the years following ($\rho$\,Pup, \citet{neiner2017}; $\beta$\,Cas, \citet{zwintz2020}; HD\,41641, \citet{thomson2020}). While recent efforts to perform large-scale studies in this field, in the hopes of generating a database of known magnetic \dsct{} stars, have had varying degrees of success \citep{thomson2023,thomson2024} the rate of detection appears to be gradually increasing over time. Regardless of the success of these studies, they inevitably provide insight and additional observational constraints, such that subsequent studies are more fine-tuned and thus more likely to provide positive results. 

Amongst the four already known magnetic \dsct{} stars from the literature, there does not appear to be any consistency in the characteristics of their magnetic fields, suggesting the existence of several regimes (e.g. weak fields versus strong fields). 
\revision{However, due to the limited size of the dataset, it is difficult to reach a conclusion on the origin of the observed diversity in both field strength and nature, and what mechanisms might be responsible.}
In order to be able to hypothesise as to the nature of this disparity, as well as assess whether any correlations exist between the stellar parameters of \dsct{} stars and the representation of magnetism in these stars, it is essential to expand this database and acquire a larger list of well-studied magnetic targets.

It is with this goal in mind that we compile a list of 13 spectropolarimetrically confirmed magnetic \dsct{} stars, that we have identified and characterised to verify their nature, and which represents all known targets of this type discovered to date. From this list, we hope to begin to visualise any possible correlations (or lack thereof) that might exist between the character of their magnetic field and key stellar parameters, and provide a first statistical image of magnetism in \dsct{} stars.

\section{Data analysis}
The targets in our sample have been compiled from the literature. We selected targets claimed to be \dsct{} stars or magnetic, or both. 
To ensure the veracity of the target characteristics in our sample, we imposed the following restrictions: i) that they have at least a single sector of photometric data from the Transiting Exoplanet Survey Satellite \citep[TESS, ][]{ricker2015} (preferably several) to verify the presence of \dsct{} pulsations, and ii) that spectropolarimetric data is available to determine the magnetic nature of the star. 

A number of young Herbig-type stars have also been shown to present both features (e.g. HD\,35929, \citealp{the1994,marconi2000}; HD\,72106, \citealp{vieira2003,wade2005}),
but for the purposes of this study we have elected to only consider targets that have evolved to the main sequence or slightly beyond.

In the following subsections, we will briefly describe the procedure performed to validate the two aforementioned selection criteria.

\subsection{TESS photometry}\label{tess}

\revision{In broad strokes, photometry from TESS was used to confirm that our targets are genuine delta Scuti pulsators and to search for low-frequency signals indicating rotational modulation. This involved using the full frame images (FFIs), as well as light curve data products from the Mikulski Archive for Space Telescopes (MAST) at the Space Telescope Science Institute\footnote{
\url{https://archive.stsci.edu/missions-and-data/tess}} when available. Our procedure is described below. }

\revision{A preliminary check of the TESS photometry for our targets clearly showed signals in the expected frequency range for delta Scuti pulsators. Following a similar procedure as described in \citet{labadie2023}, pixel-level light curves and frequency spectra were generated from the FFIs, assisted by the \textsc{Lightkurve} \citep{lightkurve}, \textsc{TESScut} \citep{brasseur2019}, \textsc{Astropy} \citep{astropy2013,astropy2018}, and \textsc{TESS_Localize} \citep{higgins2023} packages, we confirmed that all such high-frequency signals originate on-target and are not due to contamination from neighbouring sources. We also examined the frequency spectrum for the highest cadence data available (either 200s, 120s or 20s, depending on the star) to check that our initial analysis was not confused by super-Nyquist signals. The astrophysical motivation for this is that roAp stars exhibit very high frequency pulsations, typically $>$50 d$^{-1}$ \citep[e.g.][]{holdsworth2021}, which can be Nyquist reflected to lower frequencies in photometry with lower cadence. While at least one \dsct{}/-roAp hybrid is known \citep[KIC 11296437; ][]{murphy2020}, our analysis of the highest-cadence TESS data available offers no evidence for roAp pulsations in our sample. In other words, all detected high-frequency signals are fully consistent with \dsct{} pulsation. }

\revision{To search for rotational signals, we relied primarily on light curves extracted from the FFIs using simple aperture photometry (i.e. the only detrending step is the subtraction of background flux). Especially for slower rotators, detrending algorithms such as PDCSAP can distort or remove real variations \citep[e.g.][]{holdsworth2021}. Candidate rotation signals were identified for all of our targets (except for $\beta$\;Cas and $\rho$\;Pup) by the presence of a strong signal at low frequencies, usually with one or more harmonics. These rotation periods were confirmed or refined by inspecting the phase-folded photometry (after pre-whitening against the higher-frequency signals) at 0.5$\times$, 1$\times$, and 2$\times$ the candidate signal. Plots showing the rotational modulation and the pulsational frequency spectrum are given in Appendix~\ref{app:figs}. For $\rho$\;Pup, the only signal present in the low-frequency regime corresponds to one half of the TESS orbital period and is almost certainly systematic in nature – no convincing rotational signal could be found in the TESS photometry. }

Through the analysis of TESS photometry, we identify HD\,73857 as a high-amplitude \dsct{} (HADS) star. These HADS occupy a sub-region of the \dsct{} instability strip, and typically have only one or two excited radial modes (often corresponding to the fundamental mode and/or first overtone mode), with significantly larger peak-to-peak amplitudes than their more `classical' counterparts \citep{mcnamara2000,paunzen2019}. This is in agreement with the findings of previous articles \citep[e.g.][]{claret1990,ferro1994,boonyarak2011}.

From this study, we find that five stars that were already known to be magnetic (HD\,8441 and HD\,68351, \citet{babcock1958}; HD\,10783, \citet{preston1968}; HD\,81009, \citet{preston1971}; HD\,213918, \citet{cramer1980}) show the existence of \dsct{} pulsations. 

\subsection{Spectropolarimetric measurements}\label{specpol}
Datasets for our sample come from a suite of spectropolarimetric instruments, including MuSiCoS \citep{donati1999}, Narval \citep{auriere2003}, and ESPaDOnS \citep{espadons2006}. While the resulting spectra may vary slightly from instrument to instrument (e.g. wavelength range, resolution, signal-to-noise), the techniques utilised to reduce and analyse the data, as well as the interpretation of the results themselves, is identical in all cases. 

Outside of the stars studied by collaborators in previously published articles \citep[HD\,188774, $\rho$\,Pup, $\beta$\,Cas, HD\,41641, HD\,49198, HD\,36955, and HD\,63843; ][]{lampens2013,neiner2017,zwintz2020,thomson2020,thomson2024}, each new target in our sample had a variable number of sequences available for analysis, ranging from as little as a single sequence up to 42. For \revision{each spectrum}, we generated a template line mask with spectral line information retrieved from the Vienna Atomic Line Database \citep[VALD3; ][]{piskunov1995,kupka1999,ryabchikova2015}, based on values of surface gravity and effective temperature taken from the literature. These template masks were then modified to remove telluric lines, hydrogen lines, and any spectral lines that fall within the same region as hydrogen lines. The depths of individual lines were then adjusted to best fit those of the actual star, following the procedure described in \citet{grunhut2017}.

Once complete, we utilise the Least Squares Deconvolution \citep[LSD;][]{donati1997} method to analyse the spectropolarimetric data. This functions by calculating a mean spectral line shape -- using wavelength, line depth, and Landé factor as weights -- for both the Stokes I and V profiles, as well as a null profile (labelled N) used to verify that the V signal is legitimate. In the event that a target has multiple sequences performed on a given night, their profiles can be summed together, resulting in an overall improvement in the signal-to-noise ratio (S/N). The resulting LSD Stokes I and V profiles are also presented in Figs.~\ref{fig:stokes} to \ref{fig:stokes5}. We have elected not to show the N profiles, as they are all flat.

From this study, we find that one star, that was previously identified as a \dsct{} star (HD\,73857, \citet{bessell1969}) and confirmed as a HADS in this work (see above), is magnetic.

\subsection{The sample}

Considering the seven magnetic \dsct{} stars known before this work (HD\,188774, $\rho$\,Pup, $\beta$\,Cas, HD\,41641, HD\,49198, HD\,36955, and HD\,63843), the five magnetic stars found to be \dsct{} stars in Sect.~\ref{tess} (HD\,8441, HD\,68351, HD\,10783, HD\,81009, and HD\,213918), and the one \dsct{} star found to be magnetic in Sect.~\ref{specpol} (HD\,73857), we end up with a sample of 13 confirmed magnetic \dsct{} stars. This work thus almost doubles the list of such stars.

\section{Characterisation of the sample stars}

\subsection{Stellar parameters}

Having access to this hitherto unprecedented sample size of high fidelity magnetic \dsct{} stars allows us to perform a preliminary analysis into the state of magnetism and the general pulsation properties within this group of stars. To this end, we made efforts to retrieve a maximum of stellar parameters already available from the literature. This typically corresponded to effective temperature, apparent magnitude, spectral type, and mass. In many cases however, few parameters were readily available, requiring us to take estimates from the \emph{Gaia} DR3 archive \citep{gaia2023}, and/or determine them ourselves via scaling relations. Temperatures retrieved from \emph{Gaia} were determined using the Extended Stellar Parametrizer for Hot Stars (ESP-HS), which provides corrections to the stellar parameters determined for stars with effective temperatures $\gtrsim 7,500$ K, which have been shown to be less reliable when determined using the standard procedures \citep{gaia2023c}.

For luminosity, we used the parallax values from \emph{Gaia}, in combination with Eq.~\ref{eq:abs-mag} below to acquire values of absolute magnitude. We then inserted the values into Eq.~\ref{eq:luminosity}, to convert them into solar luminosities. Although extinction and bolometric corrections can impact luminosity estimates, applying them uniformly across a diverse sample is often non-trivial. We therefore adopt this simplified approach, which provides a reasonable first-order estimate sufficient for population-level analysis and comparative studies \citep[e.g. ][]{andrae2018}. \revision{These parameter relations} also assume that our stars have masses between 2 and 55 M$_\odot$ \revision{\citep{eddington1924,duric2004,eker2015}}, which is consistent with the majority of our sample and the fact that \dsct{} are intermediate mass stars. Using the same assumption, we can also calculate masses for the few targets without values in the literature using the mass-luminosity relation shown in Eq.~\ref{eq:mass}.

\begin{align}
    M_v &= m_v + 5(\log_{10} p + 1) \label{eq:abs-mag}\\ 
    \log \left(\frac{L}{L_\odot}\right) &= 0.4(4.85-M_v) \label{eq:luminosity}\\
    \frac{M}{M_\odot} & = \left(0.7\frac{L}{L_\odot}\right)^{2/7} \label{eq:mass}
\end{align}
where $M_v$ and $m_v$ correspond to the visible absolute and apparent magnitudes respectively, $p$ is the parallax, $L$ the luminosity, and $M$ the mass, the latter two both normalised with respect to solar values. 

All the stellar parameters utilised in our study are presented in Table~\ref{tab:params}, with parameters taken from the literature where possible, with those that have been inferred from \emph{Gaia} or determined via scaling relations represented by a $^\dagger$ symbol.

\subsection{Evolutionary stage}
\label{sec:cesam}

With now access to the beginnings of a statistically significant sample we want to determine stellar parameters, such as age and mass, of the targets and how they relate to each other. To this end, we have used the stellar evolution code {\sc cesam2k20} developed successively by \citet{Morel1997}, \citet{Morel2008}, \citet{Marques2013} and Manchon et al. (2025 subm.), in order to compute the stellar model grids studied. The masses range between 1.4 and 4 M$_{\odot}$, at solar metallicity. We adopted the {\sc AAG21} solar metal mixture \citep{Asplund2021} and corresponding opacity tables obtained with {\sc opal} opacities \citep{Rogers1992, Iglesias1996}, completed at low temperature with \citet{Ferguson2005} opacity tables. Conductive opacities were computed according to \citet{Potekhin1999}. Interpolation in these tables have been performed following \citet{Cassisi2021}. The {\sc opal} equation of state \citep{Rogers2002} has been used, as well as {\sc nacre} nuclear reaction rates of \citet{Aikawa2006} except for the $^{14}N + p$ reaction, for which we used the reaction rates given in \cite{Broggini2018}. The Schwarzschild criterion was used to determine convective instability, but we have tested that the Ledoux criterion does not impact the evolutionary tracks. Convection was treated using the mixing-length theory (MLT) formalism \revision{\citep{Bohm-Vitense1958,Henyey1965}} with a parameter $\alpha_{\rm MLT}=1.642$, calibrated to the Sun. The atmosphere is matched to a $T(\tau)$ law at an optical depth of $\tau=20$ \revision{\citep{Hubeny2014}}. \revision{Among several prescriptions possible in {\sc cesam2k20}, we have elected to choose diffusive overshooting as parametrized in \citet{herwig2000}, with a diffusive coefficient equivalent to an $\alpha_{ov}$. The values for the $\alpha_{ov}$ are taken following \citet{Claret2016} ($\alpha_{ov} = \alpha_{CT}$), which propose an empirical calibration varying with the stellar mass.}

For all the stars in the sample, stellar ages were approximated by interpolating between the calculated {\sc cesam2k20} evolutionary tracks, using a 10,000-by-10,000 pixel mesh grid, and taking into account the precision on our determined values of effective temperature and luminosity. Ages derived from these models are presented in column 9 of Table~\ref{tab:params}. Due to the fact that quite a number of the stars in our sample are chemically peculiar (CP), we recognise that this is not an accurate determination of age for the specific targets and \emph{should not} be used as reference for future works. Nevertheless, 
it still might provide some insight into how magnetic fields might evolve over the course of stellar lifetimes in \dsct{} stars. 

Recent studies of stellar evolution suggest that the pre-main-sequence (PMS) phase of \dsct{} star evolution overlaps those of the main-sequence (MS) or post-MS  \citep{zwintz2014,steindl2021}. At face value, this could put into question our assessment of stellar ages and evolutionary stages for our sample, particularly for the youngest/least evolved amongst them. To determine whether the stars in our sample have at least evolved to the MS (if not beyond), we assessed several criteria including infrared excess in their spectral energy density (SED) profiles, whether they were located in or near a star-forming region, and whether emission features were observed in their spectropolarimetric datasets. 
We utilised the virtual observatory SED analyser \citep[VOSA;][]{bayo2008} to identify the presence of IR excess, which compiles photometric datasets from a wide variety of sources. No IR excess was detected for any of the stars presented here, suggesting they are not PMS stars containing cool, dusty, disk material in their environment. 
Similarly, by assessing the unnormalised spectropolarimetric spectra, particularly in regions around hydrogen lines where emission features are most likely to be visible, we did not observe emission that would indicate the presence of a disk or nearby material that would indicate the stars are still in their PMS phase.
Finally, we checked whether any of the stars in our sample were attributed to star-forming regions. We found nothing in the literature, and a relative distance measurement of nearby stars for each source, using parallaxes and distances from \emph{Gaia}, did not reveal any conclusive results. 

\subsection{Magnetic analysis}

\subsubsection{Longitudinal magnetic field calculation}
From the aforementioned LSD Stokes profiles, we are able to determine the strength of the magnetic field along the line of sight \citep{rees1979,wade2000}. These longitudinal magnetic field values are calculated by defining a region around the centroid of the line in such a way as to include the full line profile while simultaneously minimising the contributions from the continuum. We can then perform the integral of the Stokes V profile in this region  which is subsequently normalised by the Stokes I profile and several constants. These $B_l$ values are presented in Table~\ref{tab:specpol}, along with $N_l$ values which are calculated in the same way but on the N profiles instead. We complete this step for each night of observation performed for a given target. 

As is commonplace with this method of determining the strength of $B_l$, we utilise a False Alarm Probability (FAP) algorithm to verify the validity of the signal in V, by comparing it to the N profile in the same region, and check that there is not any possible signals hidden in the continuum. This algorithm follows the criteria described in \citet{donati1992}, with a definite detection (DD) defined as $\text{FAP} \lesssim 10^{-5}$, a marginal detection (MD) corresponding to $10^{-5} \lesssim \text{FAP} \lesssim 10^{-3}$, and a non-detection (ND) being returned for values $\text{FAP} \gtrsim 10^{-3}$. The results of these tests have also been included in Table~\ref{tab:specpol}.

\subsubsection{Polar magnetic field strength approximations}
\label{sec:bpol}

Using these $B_l$ values, we can approximate the polar field strength $B_{\rm pol}$, which gives us an idea of the overall strength of the magnetic field at the surface of the star, assuming a dipolar field structure. This is a relatively safe assumption, as the majority of stars with strong, global magnetic fields, show dipolar structure \citep{braithwaite2004,auriere2007}. Following the method described in previous articles \citep[e.g.][]{freour2022,thomson2020,thomson2024}, we utilise an Oblique Rotator Model, first described in \citet{stibbs1950}, to calculate a value for $i$, the inclination of the rotation axis with respect to the line of sight, as follows:

\vspace{-5pt}
\begin{align}
    i &= \sin^{-1}\left(\frac{v \sin i \, P_{\rm rot}}{2\pi R}\right) \label{eq:inclination} \\
    \frac{R}{R_\odot} &= \left(\frac{T_{\text{eff}}}{T_{\text{eff},\odot}}\right)^{-2} \left(\frac{L}{L_\odot}\right)^{1/2} \label{eq:radius}
\end{align}
where $P_{\text{rot}}$ is the rotation period, $v$ is the rotational velocity along the stellar equator, $L$ has been determined using Eq.~\ref{eq:luminosity} above, and $T_{\text{eff}}$ and $R$ are the effective temperature and radius respectively, both normalised to solar values. \revision{These equations assume a spherical geometry, which may begin to break down at higher rotational velocities. While we do not expect this to cause issues for the sample presented herein, it is important to note nonetheless.}

The rotation period $P_{\rm rot}$ has been determined from TESS photometry by identifying rotational modulation, i.e. a frequency peak with harmonics in Fourier space. $v\sin i$ has been calculated from the shape of the LSD Stokes I profiles via the Fourier method, and should also represent the most reliable values for this parameter currently available. Other stellar parameters have either been taken from the literature, or retrieved from \emph{Gaia} \citep{gaia2016,gaia2023} and inserted into Eqs. \ref{eq:luminosity} and \ref{eq:radius} above in order to determine a value for $R$. We can then use this value of $i$ to determine a value for $\beta$, the obliquity angle between the rotation axis and the magnetic field axis, using:

\vspace{-5pt}
\begin{equation}
    \frac{B_l^-}{B_l^+} = \frac{\cos(i+\beta)}{\cos(i-\beta)}
\end{equation}

Finally, we utilise the equation defined in \citet{schwarzschild1950} to acquire a value for $B_{\rm pol}$:

\vspace{-5pt}
\begin{equation}
B_{\rm pol} = \frac{4(15-5u)}{15+u} \frac{B_{l}^{\pm}}{\cos(i \mp \beta)}
\label{eq:bpol}
\end{equation}
where $B_l^+ / B_l^-$ correspond to the maximum/minimum amplitude of a dipolar fit to the calculated $B_l$ values, and $u$ corresponds to the limb-darkening coefficient for a given star, retrieved from \citet{claret2019}. 

In the cases where we only have access to a couple of spectropolarimetric observations and a dipolar fit to the  $B_l$ values cannot be performed, we can nevertheless determine a lower bound of the theoretical $B_{\rm pol}$, and make the following assumptions with this in mind. First, we set $\cos(i \mp \beta) = 1$, and second we select the largest absolute value of $B_l$ and assume it is the maximum amplitude of an eventual dipolar fit. Consequentially, if we were to acquire additional datasets for these targets, the resulting $B_{\rm pol}$ can only increase, and these lower limit $B_{\rm pol}$ values have thus been represented by a '$>$' symbol in column 16 of Table~\ref{tab:specpol} accordingly.  

\subsubsection{Updates to HD\,41641}
Since the publication of \citet[][hereafter T20]{thomson2020}, we have gained access to an additional 465 spectropolarimetric observations taken with ESPaDOnS, split over 16 nights between 19 October 2023 and 15 January 2024. We adopted the same personalised line mask as for the 2016 data in T20, and followed the standard LSD procedure, including averaging together the profiles for a given night to improve the S/N. The full list of observations for HD\,41641 are visible in Table~\ref{tab:specpol}. 

\begin{figure}
    \centering
    \includegraphics[width=0.99\columnwidth]{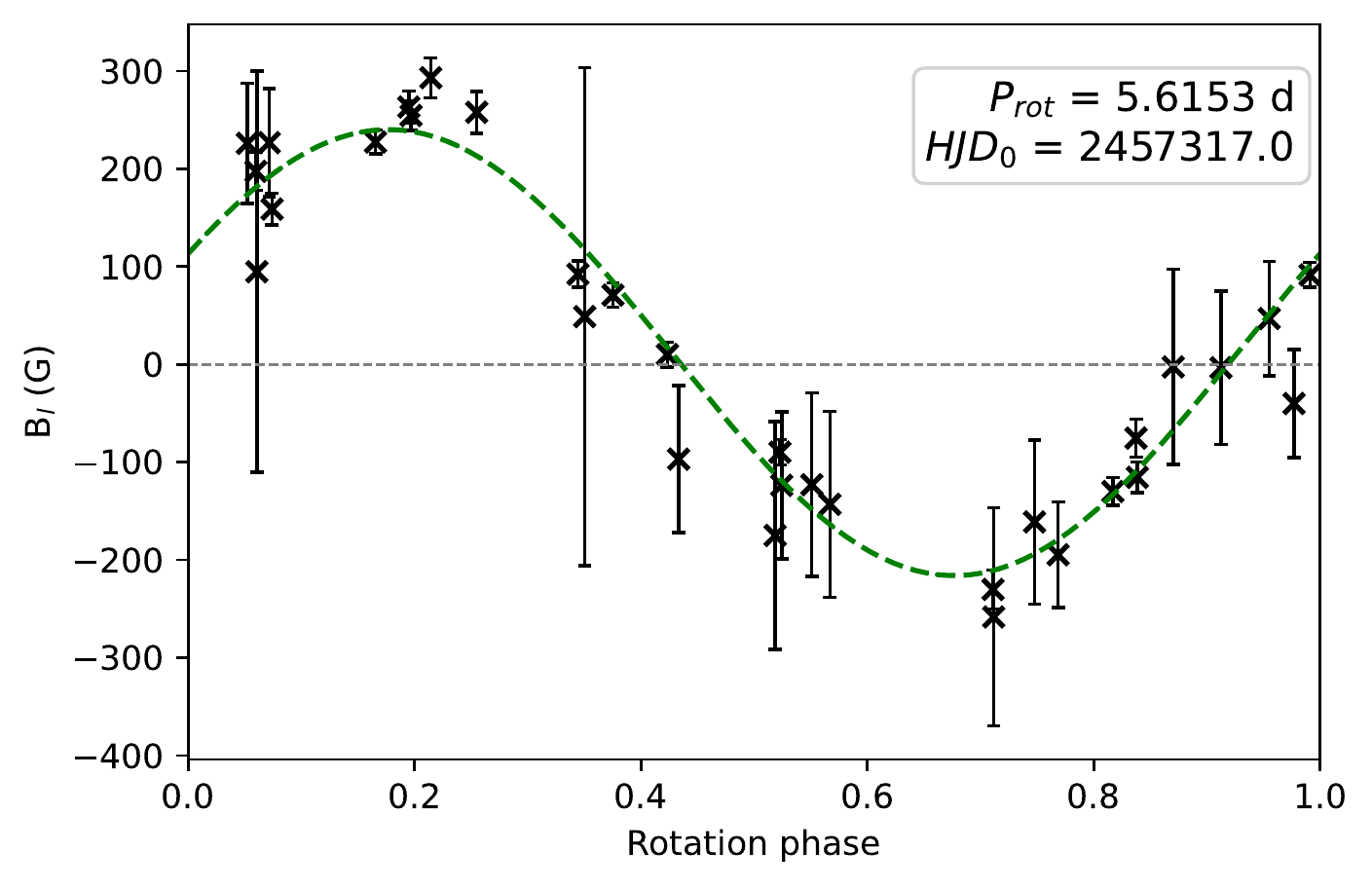}
    \caption{Calculated $B_l$ values with respect to phase, using the newly determined rotation frequency $f_{\rm rot} = 0.178085$ d$^{-1}$ for HD\,41641. Green dashed line represents the dipolar field model fit to the $B_l$ values.}
    \label{fig:hd41641_bl}
\end{figure}

\revision{A previous study of HD\,41641 by \citet[][hereafter E16]{escorza2016} revealed two harmonic frequencies (corresponding to periods of $P\sim2.8$ and $P\sim5.6$ days respectively), either of which could correspond to the rotation frequency, with E16 suggesting that the one corresponding to $P\sim5.6$~d might be the correct one. In T20, the authors determined a rotation period of $P = 2.80898876$ d, characterised by the best fitting sine wave to the $B_l$ measurements available at the time. Following the same method, and now having access to additional datasets, we determined that the rotation period was in fact closer to $P = 5.6153$ d, double the previous value and in agreement with the value originally determined by E16.} This makes a significant change to the characterisation of the field, and rather than the complex fossil field that was proposed in T20, we instead find the classical, stable, dipolar fossil field shown in Fig.~\ref{fig:hd41641_bl}. This also led to updated values for the inclination ($i = 76 \pm 12^\circ$), the obliquity ($\beta = -103 \pm 13^\circ$), and the polar magnetic field strength ($B_{\rm pol} = 748 \pm 28 \;G$), all calculated in the same way as described in Sect.~\ref{sec:bpol} above. This revised value for $B_{\rm pol}$ has also been reported in all tables and figures in this article. 

These results underscore the necessity of acquiring multiple high-quality spectropolarimetric observations to reliably constrain the properties of a star and its magnetic field. Relying on a single or limited number of measurements can lead to inaccurate or incomplete characterisation, potentially introducing misleading interpretations into the literature. The authors try to embody this notion when providing values and their respective errors for inferred/calculated parameters.

\begin{figure*}
    \centering
    \includegraphics[width=0.9\textwidth]{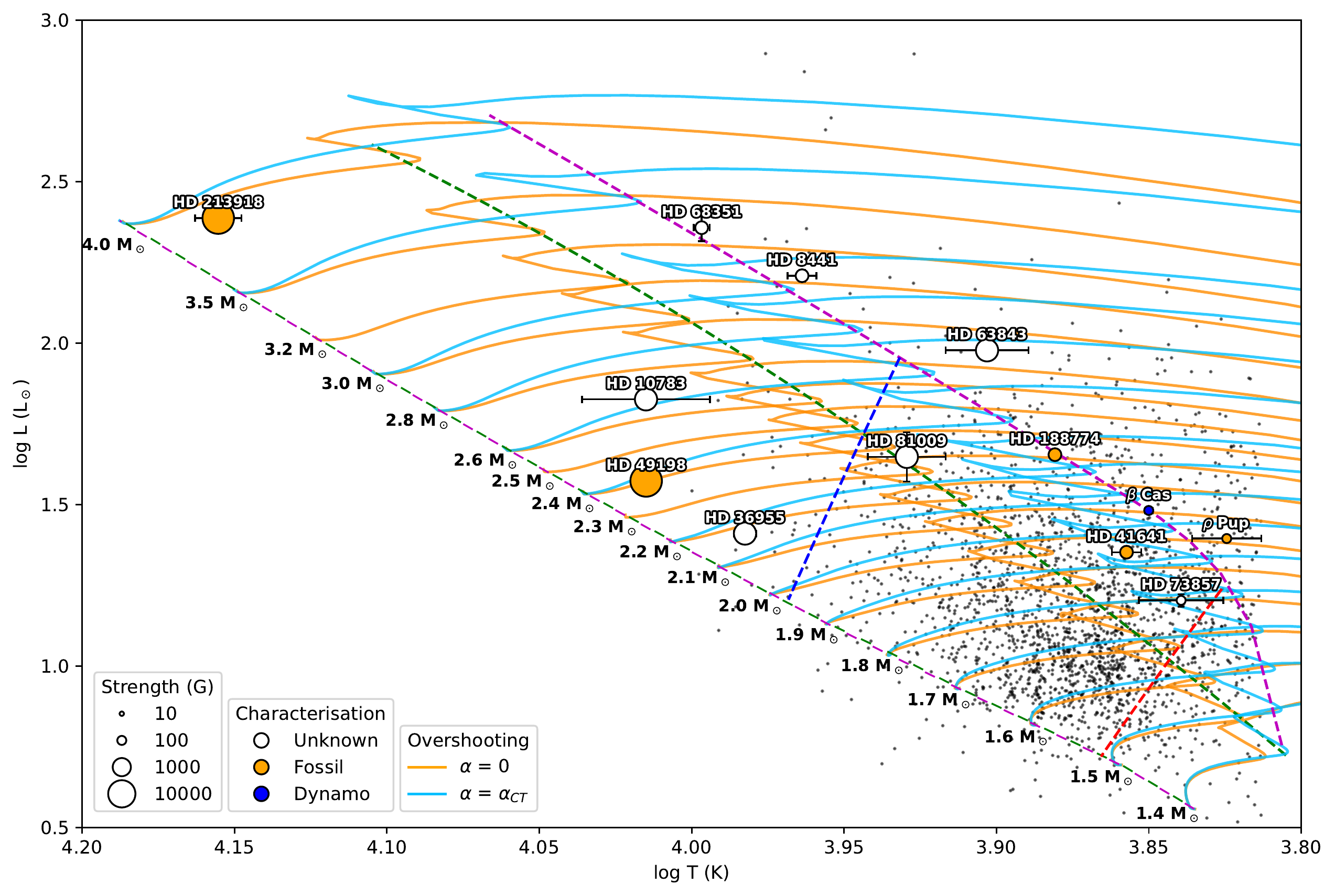}
    \caption{Representation of the 13 confirmed magnetic \dsct{} stars known to date, categorised by magnetic field characterisation (colour) and field strength (symbol size). Orange lines correspond to {\sc cesam2k20} evolutionary tracks (Manchon et al. 2025 subm.) for the mass range typical of $\delta$\,Scuti stars with no above core overshooting ($\alpha = 0$) and no magnetism included. Blue lines are the same, except with overshooting included following the \citet{Claret2016} prescription ($\alpha = \alpha_{CT}$). The approximate start and end of the main sequence are represented with green and magenta dashed lines respectively for the two types of tracks. The red and blue dashed lines correspond to the observationally constrained red- and blue-edge of the \dsct{} instability region. Black points are non-magnetic reference stars from \citet{murphy2019} classified as \dsct{} stars.}
    \label{fig:HRD}
\end{figure*}

\begin{figure*}
    \centering
    \includegraphics[width=0.9\textwidth]{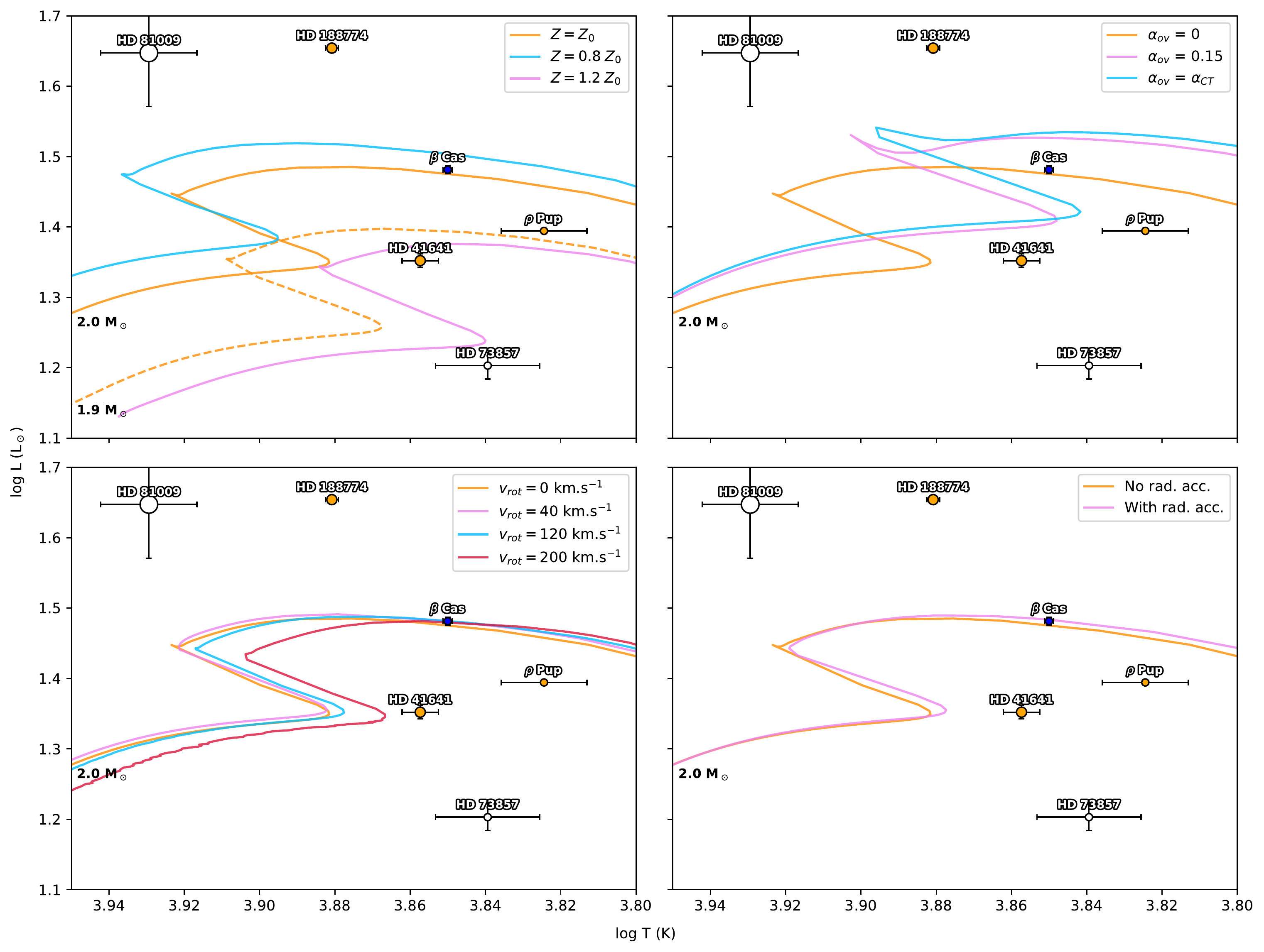}
    \caption{Variation of the 2 $M_\odot$ star {\sc cesam2k20} models presented in Fig.~\ref{fig:HRD} based on a variety of criteria, including: i) metallicity, ii) above core convective overshooting, iii) mean surface rotation velocity during the MS, and iv) radiative acceleration \revision{(see Sect.~\ref{sec:params} for more details)}. For the metallicity case, we have elected to also include the 1.9 $M_\odot$ star {\sc cesam2k20} model as comparison.}
    \label{fig:zooms}
\end{figure*}

\begin{figure*}
    \centering
    \includegraphics[width=0.9\textwidth]{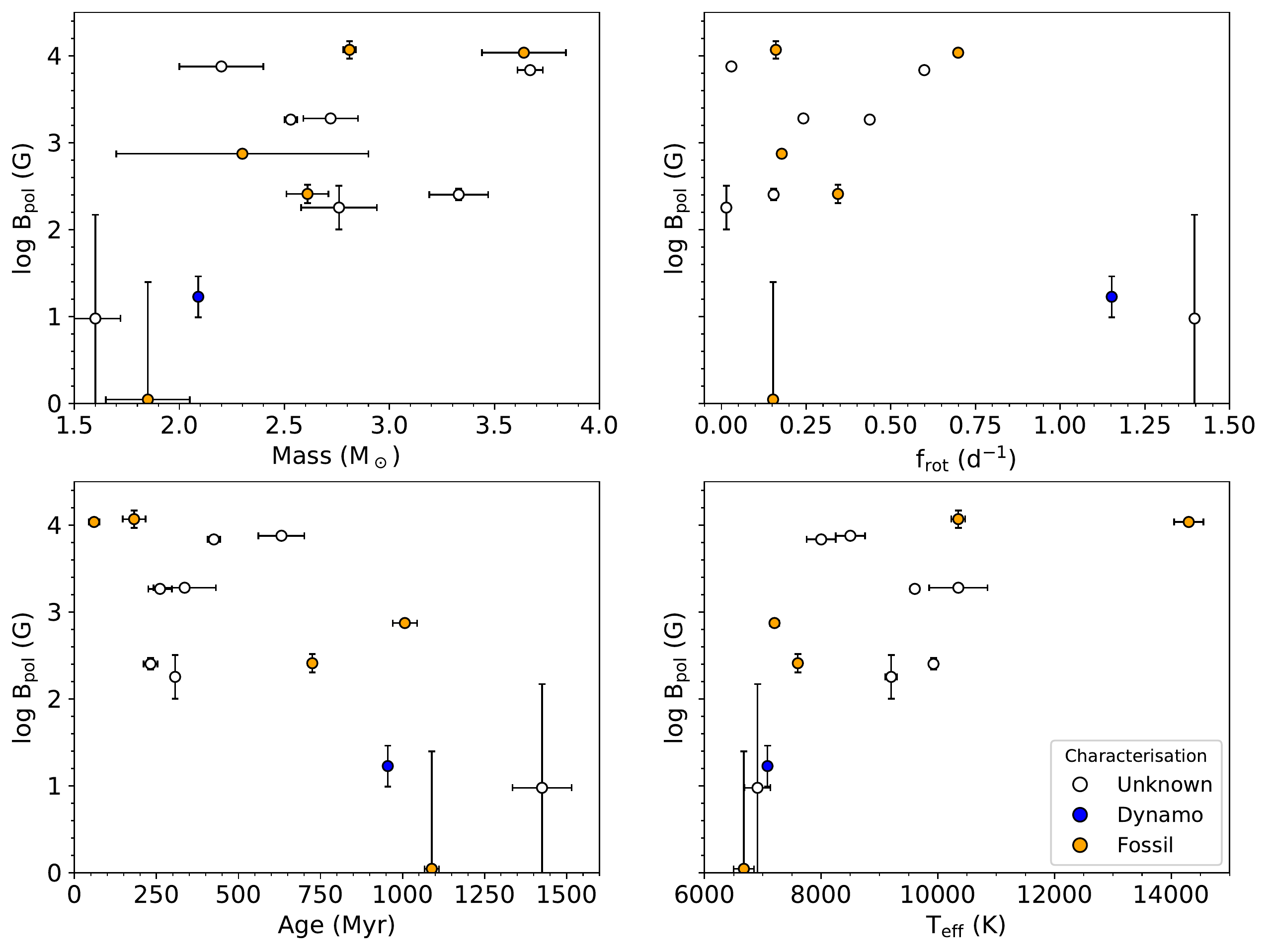}
    \caption{Comparison plots between the polar magnetic field strength and other stellar parameters for the stars in our sample. }
    \label{fig:correlations}
\end{figure*}

\begin{figure*}
    \centering
    \includegraphics[width=0.9\textwidth]{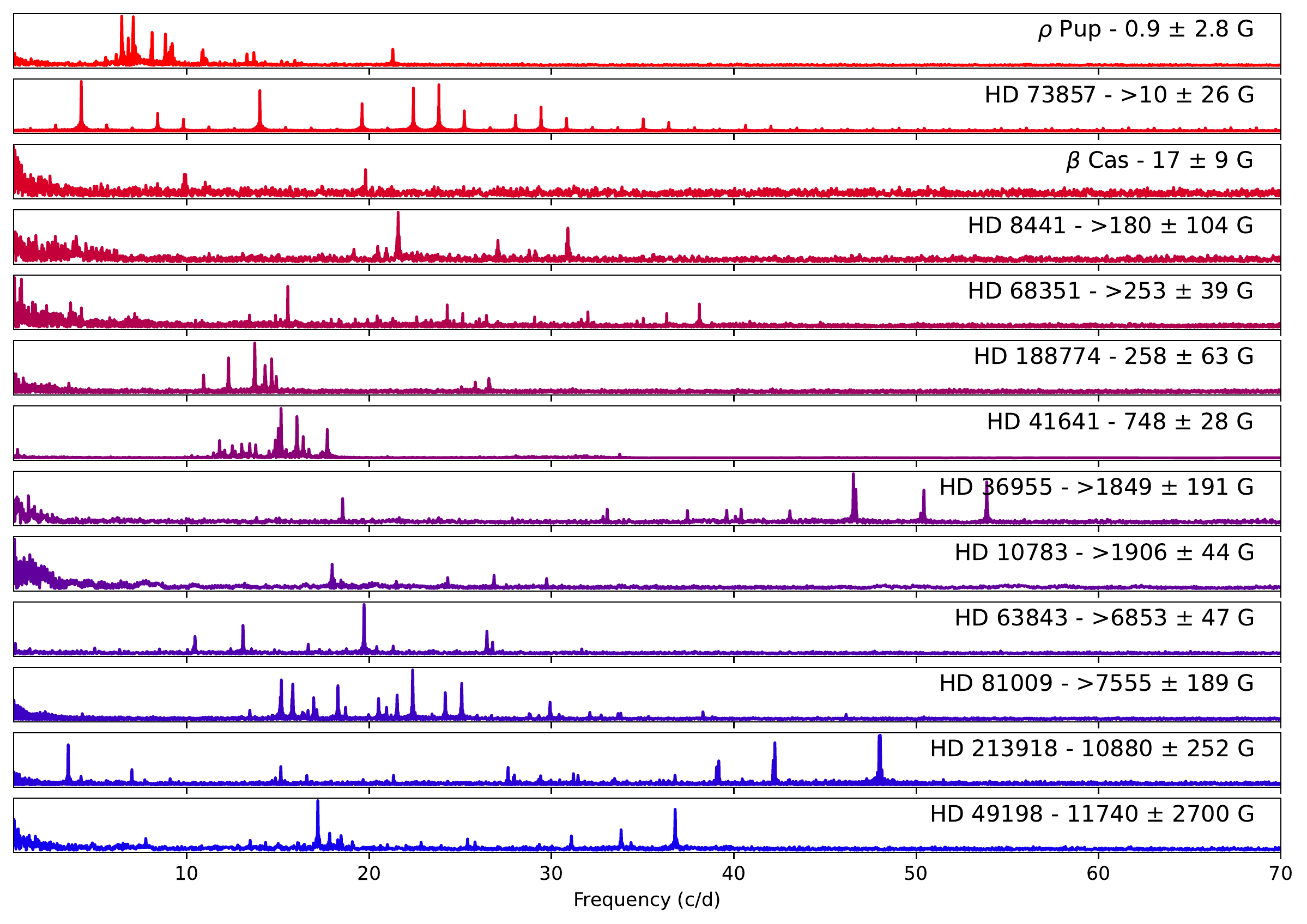}
    \caption{Lomb-Scargle periodograms for the stars in our sample, arranged in descending order with respect to the calculated polar field strength. Note that the rotational modulation has been removed to emphasise the pulsational signals.}
    \label{fig:powerspectrum}
\end{figure*}


\section{Analysis of the sample}
\subsection{Position in H-R diagram}

Using the established values for effective temperature and luminosity, we generated the Hertzsprung-Russell (H-R) diagram displayed in Fig.~\ref{fig:HRD}. To assess their approximate evolutionary stage, we used evolutionary tracks described in Sect.~\ref{sec:cesam}. 
To more easily identify the evolutionary stage, the green and magenta dashed lines visible in Fig.~\ref{fig:HRD} were generated, representing the Zero Age Main Sequence (ZAMS) and Terminal Age Main Sequence (TAMS). The ZAMS was determined from the initial values of the models, which corresponded well with a classical mass-luminosity relation for the given parameters, and the TAMS was set to determine the time at which the mass fraction of hydrogen in the core was lower than $X_c (H) =10^{-5}$.

We also retrieved the sample of A-F type stars from \citet{murphy2019} to use as reference, while also highlighting those amongst their sample that they classified as \dsct{} stars. This catalogue of reference stars has hard cutoffs at 6,500 and 10,000 K, which explains why we do not see any points on the two extremities of Fig.~\ref{fig:HRD}. \citet{murphy2019} also attempts to define the red and blue edges of the \dsct{} instability strip, constrained via observations, and presenting the following temperature-luminosity relations:

\vspace{-10pt}
\begin{align}
    \log L_{\rm red} &= -8.11\times 10^{-4} \; T_{\rm eff} + 6.672 \\
    \log L_{\rm blue} &= -1.00\times 10^{-3}\; T_{\rm eff} + 10.500
\end{align}
where $L_{\rm red}$ and $L_{\rm blue}$ are the luminosities at the red and blue edges respectively. These edges have been represented in Fig.~\ref{fig:HRD} by the red and blue dashed lines respectively.

\subsection{Distribution of sample}

As seen in Fig.~\ref{fig:HRD}, our sample spans a large area of the H-R diagram, covering a range of stellar masses and ages, which in turn assists in our attempts to study the global picture of magnetism within the family of \dsct{} stars. This includes stars on the main sequence and stars evolving beyond. 

Our sample extends well beyond the blue-edge of the \dsct{} instability strip, as determined via observations from \citet{murphy2019}. Indeed, our sample includes numerous early-A types, as well as a B6 (HD\,213918), all of which are on the hotter edge for the temperatures typically observed in \dsct{} stars. On the red-edge of this sample, we see an apparent dearth of magnetic candidates with luminosities lower than $\log L\sim 1.3 \; L_\odot$. This appears to conflict somewhat with the representation of \dsct{} stars from \citet{murphy2019}, where the majority of their reference stars are dimmer and cooler. Additionally, the few stars closest to the red edge present comparatively weak fields ($\rho$\,Pup, $\beta$\,Cas, and HD\,41641), with some presenting complexity that would be consistent with local dynamo fields often found in lower mass stars. 

This could suggest that \dsct{} stars with strong ($\gtrsim 1$ kG) magnetic fields are typically located at the hotter edge of the \dsct{} instability strip. That is to say, that they are amongst the most massive and most luminous \emph{p}-mode pulsators in the typical A-F spectral range. One should consider the fact that we are observationally biased towards hotter stars, as they are the most likely to present strong (likely fossil) magnetic fields that are relatively easy to detect, as opposed to the weak and ultra-weak fields we sometimes observe in cooler stars (e.g. $\rho$\,Pup).

That being said, we would expect stars that present weak, dynamo-generated, fields to be located in the region pertaining to stars with the lowest masses, as their internal structure more closely resembles that of low-mass stars, i.e. with convective envelopes or at least large regions associated with partial ionisation layers of hydrogen and helium. Looking at the scale of magnetic field strengths in our sample, this appears to be consistent, as the majority of the lower mass stars presented herein have globally weaker fields.

We also observe that a significant portion of our sample are more evolved stars, located at or near the TAMS, or at the start of the subgiant branch. How far beyond this point the stars are located is dependent on a number of physical factors, such as metallicity, rotation, and most importantly core overshooting. 

Due to conservation of magnetic flux, we expect more evolved stars to present weaker fields on average than those on the main sequence: while the magnetic field strength remains roughly the same, the envelope cools and expands, and as such the strength of the magnetic field we can detect at the surface decreases significantly \citep{keszthelyi2019}. An additional magnetic field decay has also been observed \citep{bagnulo2006,landstreet2007,landstreet2008}. As with the previous point, this theory appears to also hold true in our sample (see also Fig.~\ref{fig:correlations}), save for the case of HD\,63843 which is uncharacteristically strong in comparison. One thing to note is that the {\sc cesam2k20} tracks we have plotted in Fig.~\ref{fig:HRD} do not take into account magnetic fields. While the full impact of magnetic fields on stellar evolution is not entirely clear, some studies suggest that magnetism in radiative layers decreases a star's main-sequence lifetime by impacting rotation rates, chemical mixing and stellar winds \citep{maeder2005,meynet2011,keszthelyi2020}. As a result, a star containing a magnetic field would be more evolved than one without a field at the same age, primarily due to less mixing between the core and envelope.

\subsection{Variation of input physics}\label{sec:params}

To test the impact of the input physics of our {\sc cesam2k20} models on the relative positions of the stars in our sample with respect to the evolutionary tracks, we generated a range of models where we varied one parameter at a time, \revision{namely i) metallicity, ii)
above core convective overshooting, iii) mean surface rotation velocity during the MS, and iv) radiative acceleration}. These are described below, with the results of this analysis for the case of a $2 M_\odot$ model shown in Fig.~\ref{fig:zooms}.

In addition to the models calculated with solar metallicity, we generated a further two models at $\pm 0.2$ solar metallicity ($Z_0$) respectively. We note that a large portion of our sample are characterised as chemically peculiar stars, and although these models cannot reproduce the specific abundance patterns of these stars, they demonstrate how modest changes in metallicity affect the location of the evolutionary tracks, underscoring the role of metallicity in interpreting stellar properties. 

Two cases have been explored concerning overshooting above the convective core: no overshoot, or instantaneous step overshoot. For the latter, we have treated it as a further two separate cases: a constant overshoot with a parameter $\alpha_{ov}=0.15$, and an overshoot parameter which varies with the stellar mass ($\alpha_{ov}=\alpha_{CT}$), according to the prescription of \citet{Claret2016}. The latter was calculated for the full grid of models, and is also presented in Fig.~\ref{fig:HRD}.

Furthermore, we have tested how changing the angular momentum content of the star, and therefore its surface rotation velocity, impacts the main sequence lifetime. In {\sc cesam2k20} the initial angular momentum content of the star is set by initial conditions at the beginning of the pre-main-sequence. These initial conditions are fixed by considering that the star undergoes disk-locking: it is forced to co-rotate with the disk at a given rotation period $P_{\text{disk}}$, during the disk lifetime $\tau_{\text{disk}}$. Then the angular momentum distribution evolves, according to structural changes as well as transport processes following \cite{zahn1992,maeder1998} \citep[complemented by the prescriptions of ][for the turbulent viscosity coefficient]{Talon&Zahn1997, mathis2004}. \revision{Effects of the centrifugal acceleration on the stellar structure are included using the method developed in \citet{Roxburgh2004,Roxburgh2006,Manchonthesis}. The 1D solutions of structure equations computed by {\sc cesam2k20} in this case are the solutions at a characteristic latitude of $\arccos(1/\sqrt{3})$.} In this study, the two initial conditions parameters $P_{\text{disk}}$ and $\tau_{\text{disk}}$ have been tuned so that the surface rotation reaches 40, 120, \revision{and 200 km s$^{-1}$} on average during the main sequence. We observe only minor changes to both the shape of the evolutionary track and the main sequence lifetime \revision{for the 40 and 120 km s$^{-1}$ cases, which are typical for the rotation velocities observed in our sample, and a slightly more significant shift for the 200 km s$^{-1}$ case, though it too remains insufficient to explain the offset we observe. The range of rotation velocities seen across the sample remain relatively low, well below the critical rotation velocity $\Omega_{crit}$ (see column 14 of Table~\ref{tab:params})}, and thus for stars in the mass range presented in this study, this result is unsurprising.

Finally, we also calculated a model where radiative acceleration was switched on. Atomic diffusion was computed following the \citet{michaud1993} formalism and the single-valued approximation \citep{alecian2020} for radiative acceleration. Similar to the rotation case, the effect on the shape of the track varies only slightly compared to the default case.

We conclude that \revision{the instability strip corresponding to} magnetic \dsct{} stars \revision{is} shifted to hotter and more evolved \revision{regions of the H-R diagram as} compared to the non-magnetic $\delta$\,Scuti instability strip, and that the values and input physics for metallicity, overshoot, angular momentum, and atomic diffusion including radiative accelerations used in the {\sc cesam2k20} evolutionary models cannot account for this shift. However, the instability strip and the {\sc cesam2k20} models do not take magnetism into account and this might be an explanation for this discrepancy.

\subsection{\revision{Impact of critical magnetic field strength}}

We note that previous studies suggest that even a relatively weak internal magnetic field can be sufficient to enforce rigid rotation \citep[e.g.][]{spruit2002,fuller2019}, as well as effectively eliminate convective overshooting  altogether \citep[e.g.][]{briquet2012}. To test whether this is the case here, we calculated the critical magnetic field strength $B_{\rm crit}$ of each of the stars in our sample. This was performed for both the criteria as defined by Zahn \citep{zahn1992} and Spruit \citep{spruit2002} respectively, and calculated using the following equations:

\begin{align}
    B_{\rm crit,Zahn} &= \left(\frac{4\pi \rho R v_{\rm rot}}{\tau_{MS}}\right)^{1/2} \\
    B_{\rm crit,Spruit} &= \sqrt{4\pi\rho}R\left(\frac{\eta q^2 v_{\rm rot}^2}{3\pi^2 R^4}\right)^{1/3}
\end{align}
where $\rho$ is the mean density, $v_{\rm rot}$ is the rotational velocity at the equator, $\tau_{MS}$ is the age of the star since the ZAMS, $\eta$ is the diffusivity, and $q$ is the degree of differential rotation (set to 1). \revision{These two criteria are based on different assumptions, and are two different definitions of the critical magnetic field. The Zahn criterion is based on the equation of angular momentum transport and determines the critical field strength in the radiative zone needed to suppress differential rotation. On the other hand, the Spruit criterion is based on fluid dynamics and magnetic diffusion, and provides the critical initial field strength above which the magnetic field remains non-axisymmetric and rotation becomes uniform.}

As we have done previously, in the case of our sample we recovered parameter values from the literature wherever possible, and otherwise approximated them using scaling relations or models, such as those described previously in Eqs. \ref{eq:radius} and \ref{eq:inclination}. The determined values of $R$, $i$, and $B_{\rm crit,surf}$ for both criteria are visible in columns 5, 11, 17 and 18 of Table~\ref{tab:params} respectively.

For the Spruit criterion, we have elected to set the magnetic diffusivity to $\eta = 2.0\times 10^6$ cm${}^2$\,s${}^{-1}$. While this value is more representative of B-type stars \citep{augustson2011}, the diffusivity evolves as $\eta \propto r^2$, and therefore would be smaller for the majority of our sample. Consequently, \revision{decreasing the value for the diffusivity to one potentially more representative of our sample} would cause the values for $B_{\rm crit}$ to also decrease as a result.

Furthermore, for both criteria, we arbitrarily set $r=0.5\,R_*$ as the comparison point between the internal $B_{\rm crit}$ and the one at the surface $B_{\rm crit,surf}$, ensuring we consider only the radiative zone, where we expect the strength to have decreased by a factor ten following the work by \citet[][see Fig. 8 therein]{braithwaite2008}.

Combining these two aspects, it further accentuates the fact that our calculated $B_{\rm pol}$ values are almost universally larger than $B_{\rm crit,surf}$, and therefore sufficiently strong to rigidify rotation and nullify overshooting. The only three targets for which $B_{\rm pol} \lesssim B_{\rm crit,surf}$ ($\beta$\,Cas, HD\,73857, and $\rho$\,Pup) all have very weak fields, and while they are located in the \dsct{} instability region, they have evolved to the TAMS or just beyond.

\section{Theoretical discussion}  

\subsection{Excitation mechanism}

We compare in Fig.~\ref{fig:powerspectrum} our $B_{\rm pol}$ values with the frequency peaks visible in the periodograms of the stars in our sample. The propagation, damping and excitation of stellar oscillations can be affected in different ways because of the presence of a magnetic field. First, the Lorentz force can directly modify the dynamics of stellar pulsations that affects their eigenfrequencies and their velocity field \citep[e.g][]{reese2004,mathis2011,rui2024}; this is the so-called direct effect of magnetism on stellar oscillations. Second, the Lorentz force modifies the equilibrium structure of the star \citep[e.g.][]{duez2010}. This modifies the magneto-hydrostatic cavity in which stellar pulsations are propagating and thus their properties. Such effects can be effective in layers close to the stellar surface where the density strongly diminishes that allows the Lorentz force to compete with the effects of self-gravity. In deepest layers, the gravity and the gaseous pressure strongly dominate the magnetic force and pressure, respectively, and this effect becomes weak. This indirect effect can therefore be neglected for gravity/gravito-inertial modes propagating in the deep interiors of stars \citep[e.g.][]{mathis2023} while it should be taken into account for acoustic modes propagating in their outer layers \citep{gough1990}. \revision{In both these cases, we expect to see a change in the pulsation peak structure.}

Since the oscillation power spectra observed in our magnetic stars represented in Fig.~\ref{fig:powerspectrum} show a standard pattern, we consider a third way in which magnetic fields can impact pulsations\revision{, i.e. by modifying their excitation and damping rates. An example of this }is the modification of stellar turbulent convection by a magnetic field \citep[e.g.][]{gough1966,stevenson1979,hotta2018}. The Lorentz force can then diminish progressively the efficiency of convection and affects the injection of its energy into stellar pulsation modes. This has been recently predicted theoretically by \citet{bessila2024} in the case of solar-type pulsators\revision{, which is} in agreement with the lack of observed stochastically-excited pulsations in magnetically-active solar-like stars \citep{chaplin2011,mathur2019}. 

The case of \dsct{} stars is different\revision{, however}. In these classical pulsators, stellar oscillations are excited by the $\kappa$-mechanism due to the Helium ionisation, which \revision{helps} to explain the blue edge of the instability strip where pulsations are generally observed, while the properties of convection must be considered carefully to explain its red edge \citep[e.g.][]{dupret2004,dupret2005b}. The depth of the instability region is linked to the frequency range in which we can observe pulsations. As such, we would expect to see a gradual increase in the frequency range within which the \dsct{} frequencies are located, with respect to growing values of $B_{\rm pol}$. This correlation is not particularly visible in Fig.~\ref{fig:powerspectrum}, which suggests that magnetism is not the only ingredient in the determination of the position of \dsct{} frequencies. It should also be emphasised that for many of our targets, only a single spectropolarimetric dataset was available, making the calculation of $B_{\rm pol}$ correspond to a lower-bound rather than a true assessment. As such, should additional datasets become available, and the corresponding $B_{\rm pol}$ values prove to increase as a result, the order in which these stars are displayed in Fig.~\ref{fig:powerspectrum} could change and thus potentially reveal such a trend.

In the present work, we can see in Fig.~\ref{fig:HRD} that some of the observed magnetic stars are far beyond the predicted blue edge of the instability strip. These stars are more massive with the strongest observed magnetic field amplitudes, in particular for the two hosting a fossil field (HD\,49198 and HD\,213918). We can therefore focus our discussion for those stars on the $\kappa$-mechanism. As of today, no theoretical formalism exists to model this excitation mechanism in presence of magnetism. We are thus restricted to a qualitative discussion. If the magnetic field, and the associated Lorentz force, are sufficiently strong, the thermodynamical quantities such as the (magneto-) hydrostatic background density, temperature and pressure can be perturbed. This can be particularly drastic in the external layers of the star where the gas pressure and density drop \citep[e.g.][]{duez2010}. As a consequence, the values of the stellar opacity and of its derivatives as a function of these thermodynamical quantities can be modified that may lead to a modification of the position of the blue edge of the instability strip (we refer the reader to the review of \citet{samadi2015} for a detailed explanation of the formalism; see their equations 3.26, 3.27 and 3.28). Such a possible modification of the $\kappa$-mechanism by stellar magnetic fields must be quantified in the near future but is beyond the current work. Since the perturbations of the thermodynamical quantities by a magnetic field scale with the local ratio of the magnetic force to the gravity, such future predictions may allow us to constrain the magnetic field amplitude in the layers where stellar oscillations are excited.

\subsection{Stellar modelling}

Another intriguing behaviour of the observed magnetic stars is that many of them seem to be already beyond their terminal age for the main sequence. No evident reason can explain such a trend. The more plausible explanation can rely on the uncertainties we have on our stellar modelling. The stars we are studying here are relatively slow rotators when compared to the typical distribution of velocities observed in \dsct{} stars \citep[e.g.][]{wang2025}, and host either fossil or dynamo-generated fields. First, both rotation and magnetic fields are known to affect the secular evolution and mixing in stellar radiative layers 
\citep[we refer the reader to the reviews by ][]{maeder2009,mathis2013,aerts2019}. Given the challenges we face to model these dynamical processes in a realistic and robust way in stellar structure and evolution codes, predictions obtained by our state-of-the-art stellar models must be considered carefully. The same discussion can be done for the modelling of stellar convective cores and envelopes. Indeed, both rotation and magnetic fields can drastically modify the convective super-adiabaticity and characteristic velocities and length scales \citep[e.g.][]{chandrasekhar1961,gough1966,stevenson1979,barker2014,augustson2019,hotta2018,korre2021,bessila2025}. Such modifications of the properties of stellar convection and of the related over-undershooting can modify predictions for stellar structure and ages \citep[e.g.][]{ireland2018,dumont2021}. A last key point must be considered for the fastest rotating stars. For those stars, it may be necessary to adopt a full 2D modelling of their structure \citep{mombarg2023,mombarg2024} that may lead to a drastically different prediction for their age \citep{rieutord2024}.

\section{Statistical analysis}

\revision{To look for trends between stellar parameters and the global magnetic field strenth of the stars in our sample, we generated the plots shown in Fig.~\ref{fig:correlations}. Notably, we considered the potential effects of mass, rotation velocity, age, and effective temperature on the distribution of field strengths. These impacts are described briefly in the following subsections.}

\revision{In addition,} to check the impact of \revision{the} magnetic field on pulsations, we investigated whether any correlation was visible between the strength of the magnetic field and the location of the \dsct{} frequency peaks in Fourier space, generating the array of plots visible in Fig.~\ref{fig:powerspectrum}. These periodograms have been pre-whitened against lower-frequency signals, including rotation and any instrumental trends. While some low-frequency variability is still visible in a few cases, they do not hinder the identification of the \dsct{} pulsations visible within the Fourier spectra.

\subsection{Stellar mass regimes}

\revision{In the top-left panel of Fig.~\ref{fig:correlations}, we see what appears to be two regimes: the weaker fields in low-mass stars and the stronger fields in the high-mass stars. This is expected as, at} low masses, we expect to be in a regime of stars predominantly presenting weak dynamo fields, as opposed to the strong fossil fields we observe at higher masses.

\revision{There is evidence that more massive stars tend, on average, to host stronger global magnetic fields. For example, intermediate-mass Ap/Bp stars ($\approx 2-6\; M_\odot$) often show fields of several hundred Gauss to multiple kilogauss, with lower mass stars rarely displaying fields above $\sim$300 G \citep{auriere2007,power2008}. Among higher mass stars, studies show instances of stars demonstrating fields of the order of several kilogauss \citep[e.g.][]{shultz2019}. However, further studies also show that the relationship between field strength and mass is not simple — magnetic flux appears to increase with mass (consistent with larger surface area), but the field strength alone does not always show a clean monotonic trend \citep{kochukhov2006}.}

We should therefore see a `ramp-up' of the strength of polar magnetic fields in the intermediate region. Aside from the outlier at 2.2 M$_\odot$, corresponding to HD\,81009, this is essentially what we observe in our sample. 

Since stellar mass and effective temperature are intrinsically linked, we see a similar ``ramp-up" for the T$_{\rm eff}$ plot (see \revision{the bottom-right panel of Fig.~\ref{fig:correlations}, and also} Fig.~\ref{fig:ps_teff}), in that hotter stars are more likely to host stronger magnetic fields. We note however that four stars in our sample have $T_{\rm eff} \gtrsim 10,000$ K, which is unusual for \dsct{} stars. Theoretical studies have shown that rotation impacts the width and location of the $\kappa$-mechanism instability region of \revision{classical pulsating stars}, shifting it toward higher luminosities and effective temperatures \citep{dupret2005,townsend2005}, which might help to explain this offset.

\subsection{Effect of rotation}

With regards to \revision{rotation velocity}, an initial assessment \revision{from the top-right panel of Fig.~\ref{fig:correlations}} suggests a trend of faster rotators leaning towards lower $B_{\rm pol}$ values, ignoring the apparent outlier at \revision{$f_{rot}=0.15$ d$^{-1}$} which corresponds to HD\,213918. It should be noted, however, that we are biased in the $v\sin i$ values of the stars in our sample. We detect weak fields at a wide range of $v\sin i$ values, which shows that our spectropolarimetric observations achieved sufficient sensitivity to be able to detect weak fields for all our targets. We also managed to detect a number of stronger fields, but with only four such targets we are unable to conclude with regards to the distribution of these strong fields with respect to $v\sin i$\revision{ or $f_{rot}$}. Undoubtedly, additional targets are required to test to what extent this trend remains true, ideally with at least a few higher velocity examples ($v \sin i > 100$ km s$^{-1}$). 

\subsection{Distribution of field strengths}

\begin{figure}
    \centering
    \includegraphics[width=0.95\columnwidth]{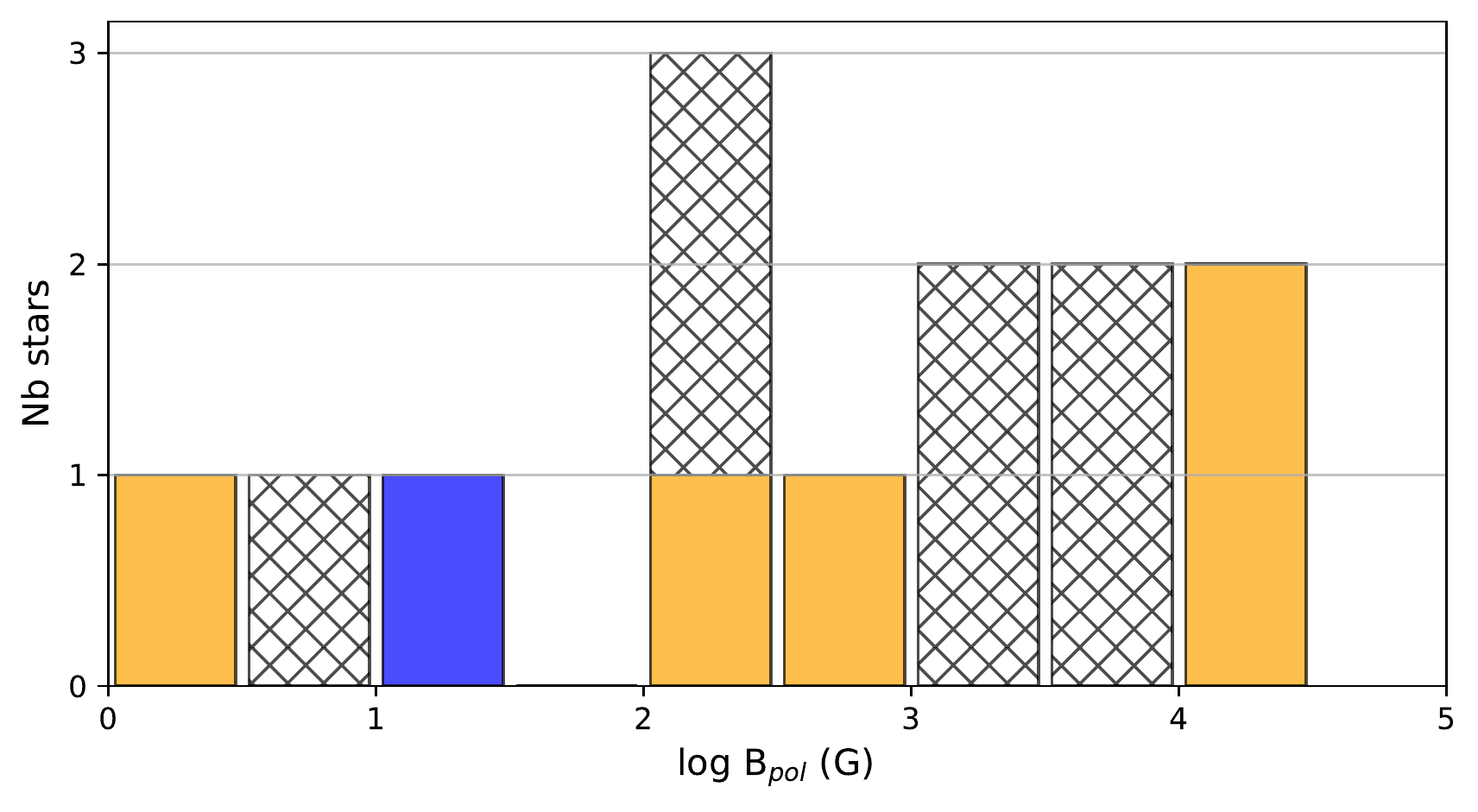}
    \caption{Histogram of polar magnetic field strengths B$_{\rm pol}$ for our sample, using the same colour classification for field characterisation as Fig.~\ref{fig:HRD}, i.e. yellow for fossil fields, blue for dynamo fields, white (hashed) for unknown field type.}
    \label{fig:hist}
\end{figure}

\revision{In Fig.~\ref{fig:hist}, we generated the distribution of field strengths in our sample, which can then be} compared with one of magnetic OBA stars from \citet[][c.f. their Figure 6]{shultz2019}. The authors observe a spread that is mostly contained between 300 G and 30 kG, with a mean value of the order of 3 kG, over a sample of $\sim150$ magnetic OBA stars. When performing a similar investigation with our sample we observe a peak of the order of a couple of hundred Gauss, though the variance between the main peak and the next largest bin is a difference of a single star and admittedly our sample size is significantly smaller. As such, this distribution is not yet statistically significant, but could suggest that globally the strengths of magnetic fields in \dsct{} stars are on average smaller than those of classical (non-variable) OBA-type stars.

In addition, we have estimated what value $B_{\rm pol}$ might take at the ZAMS for each of the stars in the sample. This can be done using the relation
\begin{equation}
    B_{\rm pol}(ZAMS) = \frac{B_{\rm pol}(t)\times R(t)^2}{R(ZAMS)^2}
\end{equation}
where $t$ corresponds to the current age of the star. This results in the values shown in column 19 of Table~\ref{tab:params}. Evidently this value scales with the age and evolutionary stage of the star, and as expected the value increases in all cases as the star gets younger. While in some cases noticeably larger, the resulting distribution (5-41,000 G) are still well within the standard range of field strengths for OBA-type stars observed by \citet{shultz2019}. 

\section{Conclusion}
We present a curated list of 13 magnetic \dsct{} stars, assembled from a variety of sources, which we identify both as \dsct{} stars and magnetic stars. This constitutes the most comprehensive list of these objects to date, with the expectation that it will continue to be expanded upon as the authors and members of the community pursue studies in this field. 

Being the start of a statistically significant database, we have sought to identify whether any trends might be observed within this sample, by considering their positions in the H-R diagram relative to each other and other `classical' stars, as well as comparing the polar magnetic field strength values calculated as part of this study with the sample's stellar parameters.

We observe that our sample is overall shifted towards higher temperatures, luminosities and thus masses, with respect to a sample of A-F type reference stars and the observationally constrained instability strip of non-magnetic \dsct{} stars. This could be due to the impact of the magnetic field on the \dsct{} pulsation excitation, shifting these pulsations to hotter stars. A significant portion of our sample also appears to be slightly evolved stars, having reached the TAMS or progressed beyond the main-sequence. Finally, stars presenting the strongest polar magnetic fields are amongst the hottest in our sample, as expected from fossil versus dynamo fields.

When comparing with stellar parameters, we observe behaviour that is generally consistent with theory, but note that there exists several observational biases in the parameter space of our sample. The latter is skewed towards targets with lower $v \sin i$ values, which reduces the required exposure time of spectropolarimetric observations, and hotter effective temperatures, which are more likely to present strong magnetic fields and thus are also easier to detect.

Finally, despite being predicted by theory, we do not see a clear correlation between surface magnetic field strength and the location of \dsct{} peaks in Fourier space. This suggests that stellar magnetism is not the sole ingredient for the differences in position and spacing of \dsct{} frequencies.

While the list of magnetic \dsct{} stars presented here remains relatively small, it is the first study of its kind for \dsct{} stars, and will hopefully pave the way for subsequent studies into the interplay between stellar magnetism and pulsating variable stars.

\begin{acknowledgements}
We thank Yves Frémat for his insight into working with Gaia data, Louis Manchon for valuable discussions regarding the {\sc cesam2k20} evolutionary tracks presented herein, and both Victoria Antoci and Dominic Bowman for insightful discussions on the content of the article.
This work is based on observations obtained at the Canada-France-Hawaii Telescope (CFHT) which is operated by the National Research Council (NRC) of Canada, the Institut National des Sciences de l'Univers of the Centre National de la Recherche Scientifique (CNRS) of France, and the University of Hawaii. This research has made use of the SIMBAD database operated at CDS, Strasbourg (France), and of NASA's Astrophysics Data System (ADS).This paper includes data collected by the TESS mission, which are publicly available from the Mikulski Archive for Space Telescopes (MAST). Funding for the TESS mission is provided by NASA's Science Mission directorate. This work has made use of data from the European Space Agency (ESA) mission Gaia (https://www.cosmos.esa.int/gaia), processed by the Gaia Data Processing and Analysis Consortium (DPAC, https://www.cosmos.esa.int/web/gaia/dpac/consortium). Funding for the DPAC has been provided by national institutions, in particular the institutions participating in the Gaia Multilateral Agreement. This publication makes use of VOSA, developed under the Spanish Virtual Observatory (SVO) project funded by MCIN/AEI/10.13039/501100011033/ through grant PID2020-112949GB-I00. VOSA has been partially updated by using funding from the European Union's Horizon 2020 Research and Innovation Programme, under Grant Agreement nº 776403 (EXOPLANETS-A). K.T.P. gratefully acknowledges UK Research and Innovation (UKRI) in the form of a Frontier Research grant under the UK government's ERC Horizon Europe funding guarantee (SYMPHONY; PI Bowman; grant number: EP/Y031059/1). S.M. acknowledges support from the European Research Council (ERC) under the Horizon Europe programme (Synergy Grant agreement 101071505: 4D-STAR), from the CNES SOHO-GOLF and PLATO grants at CEA-DAp. C.N., R.-M.O and S.M. acknowledge support from PNPS (CNRS/INSU). J.L.-B acknowledges support from the European Union (ERC, MAGNIFY, Project 101126182). While partially funded by the European Union, views and opinions expressed are however those of the author only and do not necessarily reflect those of the European Union or the European Research Council. Neither the European Union nor the granting authority can be held responsible for them.
\end{acknowledgements}

\bibliographystyle{aa}
\bibliography{bibliography}

\clearpage
\begin{appendix}

{\onecolumn
\section{Tables}
\subsection{Spectropolarimetric observations}
\begin{table}[ht!]
\caption{Summary of spectropolarimetric observations for the sample. }
\resizebox{\textwidth}{!}{%
\begin{tabular}{ccccccccccccc}
\hline
ID & TIC & Obs date & Mid HJD & Instrument & Strategy (s) & Mean $\lambda$ (nm) & Mean Lande & FAP & $B_l \pm \sigma B$ (G) & $N_l \pm \sigma N$ (G) & $B_{pol} \pm \sigma B$ (G) & Characterisation \\
\hline
$\beta$\,Cas & 396298498 & 2013-11-03 & 2456600.2627 & Narval & 1x4x65 & 528.7963 & 1.199 & ND & 0.6 ± 6.6 & -1.1 ± 6.6 & 16.9 ± 9.2 & Dynamo \\
 &  & 2014-09-24 & 2456925.6165 & Narval & 3x4x65 & 523.7686 & 1.200 & MD & 1.3 ± 2.5 & -4.6 ± 2.5 &  &  \\
 &  & 2014-12-19 & 2457011.2941 & Narval & 5x4x65 & 522.2549 & 1.200 & DD & 3.9 ± 1.9 & 0.0 ± 2.0 &  &  \\
 &  & 2014-12-21 & 2457013.2990 & Narval & 5x4x65 & 524.663 & 1.200 & DD & 0.8 ± 2.1 & 1.4 ± 2.1 &  &  \\
 &  & 2015-12-01 & 2457358.3578 & Narval & 10x4x65 & 528.0618 & 1.200 & DD & 0.6 ± 2.6 & 1.6 ± 2.6 &  &  \\
 &  & 2015-12-02 & 2457359.3592 & Narval & 10x4x65 & 526.2186 & 1.200 & MD & -1.5 ± 3.2 & 2.1 ± 3.2 &  &  \\
 &  & 2015-12-05 & 2457362.2862 & Narval & 5x4x65 & 527.7356 & 1.200 & MD & -3.5 ± 2.6 & -1.3 ± 2.6 &  &  \\
 &  & 2015-12-06 & 2457363.3513 & Narval & 15x4x65 & 524.9052 & 1.200 & DD & -0.7 ± 2.1 & -0.7 ± 2.1 &  &  \\
 &  & 2015-12-07 & 2457364.3792 & Narval & 10x4x65 & 523.9241 & 1.201 & MD & -3.5 ± 2.2 & 0.9 ± 2.2 &  &  \\
 &  & 2015-12-09 & 2457366.3073 & Narval & 5x4x65 & 535.774 & 1.200 & DD & 3.4 ± 2.6 & -3.9 ± 2.6 &  &  \\
 &  & 2015-12-11 & 2457368.3693 & Narval & 5x4x65 & 527.5769 & 1.200 & DD & 2.8 ± 2.2 & -1.8 ± 2.3 &  &  \\
 &  & 2015-12-12 & 2457369.3464 & Narval & 10x4x65 & 526.3156 & 1.201 & DD & -0.6 ± 2.6 & -1 ± 2.6 &  &  \\
 &  & 2015-12-13 & 2457370.3799 & Narval & 5x4x65 & 529.2903 & 1.200 & ND & 1.7 ± 2.4 & -2.7 ± 2.4 &  &  \\
HD 8441 & 238659021 & 2004-07-30 & 2453217.6484 & MuSiCoS & 1x4x800 & 529.8070 & 1.202 & MD & -3 ± 32 & 43 ± 32 & >180 ± 104 & Unknown \\
 &  & 2004-09-31 & 2453249.6704 & MuSiCoS & 1x4x600 & 530.7108 & 1.201 & DD & 56 ± 33 & -53 ± 33 &  &  \\
HD 10783 & 257921991 & 2018-12-23 & 2458476.7128 & ESPaDOnS & 1x4x168 & 531.4009 & 1.198 & DD & -574 ± 13 & -3 ± 9 & >1906 ± 44 & Unknown \\
 &  & 2018-12-24 & 2458477.7373 & ESPaDOnS & 1x4x168 & 531.9343 & 1.198 & DD & 458 ± 25 & 0 ± 9 &  &  \\
HD 36955 & 427377135 & 2021-09-01 & 2459460.0706 & ESPaDOnS & 1x4x656 & 526.7640 & 1.208 & DD & -581 ± 60 & 35 ± 58 & >1849 ± 191 & Unknown \\
HD 41641 & 436428531 & 2015-10-21 & 2457317.6957 & Narval & 8x4x27 & 532.0358 & 1.188 & ND & 438 ± 145 & 330 ± 145 & 748 ± 28 & Dipolar \\
 &  & 2015-11-09 & 2457336.7578 & Narval & 3x4x27 & 535.7854 & 1.191 & ND & -175 ± 117 & -100 ± 117 &  &  \\
 &  & 2015-11-11 & 2457338.7326 & Narval & 5x4x27 & 533.5284 & 1.190 & ND & -3 ± 100 & -29 ± 100 &  &  \\
 &  & 2015-11-12 & 2457339.7545 & Narval & 8x4x27 & 531.8484 & 1.190 & DD & 226 ± 61 & 40 ± 61 &  &  \\
 &  & 2015-11-16 & 2457343.6598 & Narval & 8x4x27 & 530.9734 & 1.190 & ND & -161 ± 84 & -16 ± 84 &  &  \\
 &  & 2016-03-15 & 2457463.3385 & Narval & 8x4x27 & 557.8794 & 1.186 & ND & 95 ± 205 & -85 ± 205 &  &  \\
 &  & 2016-03-20 & 2457468.3606 & Narval & 7x4x27 & 530.4079 & 1.192 & ND & 47 ± 58 & 19 ± 58 &  &  \\
 &  & 2016-10-08 & 2457670.6344 & Narval & 10x4x27 & 528.7043 & 1.190 & DD & -40 ± 55 & -6 ± 55 &  &  \\
 &  & 2016-10-27 & 2457689.5768 & Narval & 10x4x27 & 545.8406 & 1.185 & DD & 49 ± 255 & -105 ± 255 &  &  \\
 &  & 2016-10-28 & 2457690.7036 & Narval & 10x4x27 & 541.0206 & 1.188 & DD & -123 ± 94 & -45 ± 94 &  &  \\
 &  & 2016-10-29 & 2457691.6058 & Narval & 10x4x27 & 537.2884 & 1.189 & MD & -258 ± 112 & 27 ± 112 &  &  \\
 &  & 2016-10-30 & 2457692.7331 & Narval & 10x4x27 & 536.8760 & 1.189 & DD & -3 ± 79 & -85 ± 79 &  &  \\
 &  & 2016-10-31 & 2457693.6274 & Narval & 10x4x27 & 527.2550 & 1.191 & DD & 227 ± 55 & 73 ± 55 &  &  \\
 &  & 2016-11-02 & 2457695.6581 & Narval & 10x4x27 & 528.4211 & 1.190 & DD & -97 ± 75 & -146 ± 75 &  &  \\
 &  & 2016-11-25 & 2457718.6306 & Narval & 10x4x27 & 542.8027 & 1.188 & DD & -124 ± 75 & -6 ± 75 &  &  \\
 &  & 2016-12-01 & 2457724.4852 & Narval & 10x4x27 & 537.2719 & 1.189 & DD & -143 ± 95 & 72 ± 95 &  &  \\
 &  & 2016-12-02 & 2457725.6170 & Narval & 10x4x27 & 524.5040 & 1.192 & DD & -195 ± 54 & -30 ± 54 &  &  \\
 &  & 2023-10-19 & 2460238.0503 & ESPaDOnS & 30x4x21 & 521.7209 & 1.195 & DD & 263 ± 16 & 28 ± 16 &  &  \\
 &  & 2023-10-20 & 2460239.0633 & ESPaDOnS & 30x4x22 & 519.1949 & 1.196 & DD & 71 ± 12 & -4 ± 12 &  &  \\
 &  & 2023-10-24 & 2460242.9876 & ESPaDOnS & 30x4x23 & 527.5646 & 1.195 & DD & 159 ± 16 & 7 ± 16 &  &  \\
 &  & 2023-10-25 & 2460244.0024 & ESPaDOnS & 15x4x24 & 518.926 & 1.196 & DD & 258 ± 21 & -2 ± 21 &  &  \\
 &  & 2023-12-01 & 2460280.9651 & ESPaDOnS & 30x4x25 & 524.8393 & 1.194 & DD & -75 ± 19 & -58 ± 19 &  &  \\
 &  & 2023-12-03 & 2460282.9843 & ESPaDOnS & 30x4x26 & 522.4993 & 1.195 & DD & 255 ± 15 & 18 ± 15 &  &  \\
 &  & 2023-12-29 & 2460308.9267 & ESPaDOnS & 30x4x27 & 521.7346 & 1.195 & DD & -130 ± 14 & 0 ± 14 &  &  \\
 &  & 2023-12-30 & 2460309.9050 & ESPaDOnS & 30x4x28 & 521.2746 & 1.195 & DD & 92 ± 13 & 2 ± 13 &  &  \\
 &  & 2023-12-31 & 2460310.8843 & ESPaDOnS & 45x4x29 & 518.6848 & 1.196 & DD & 228 ± 12 & -3 ± 12 &  &  \\
 &  & 2024-01-01 & 2460311.8886 & ESPaDOnS & 30x4x30 & 520.0929 & 1.196 & DD & 92 ± 13 & -30 ± 13 &  &  \\
 &  & 2024-01-02 & 2460312.8903 & ESPaDOnS & 30x4x31 & 520.8373 & 1.195 & DD & -90 ± 13 & -16 ± 13 &  &  \\
 &  & 2024-01-03 & 2460313.9474 & ESPaDOnS & 30x4x32 & 531.2326 & 1.193 & DD & -230 ± 20 & 24 ± 20 &  &  \\
 &  & 2024-01-05 & 2460315.9062 & ESPaDOnS & 30x4x33 & 529.1782 & 1.194 & DD & 198 ± 20 & 14 ± 20 &  &  \\
 &  & 2024-01-06 & 2460316.7739 & ESPaDOnS & 15x4x34 & 519.3949 & 1.196 & DD & 293 ± 20 & -11 ± 20 &  &  \\
 &  & 2024-01-07 & 2460317.9476 & ESPaDOnS & 30x4x35 & 518.9205 & 1.196 & DD & 10 ± 13 & 15 ± 13 &  &  \\
 &  & 2024-01-15 & 2460325.8939 & ESPaDOnS & 30x4x36 & 526.3133 & 1.195 & DD & -116 ± 16 & -8 ± 16 &  &  \\
HD 49198 & 16485771 & 2021-11-26 & 2459545.8924 & ESPaDOnS & 1x4x319 & 525.9691 & 1.206 & DD & -1065 ± 59 & -56 ± 50 & 11740 ± 2700 & Dipolar \\
 &  & 2021-11-27 & 2459547.0635 & ESPaDOnS & 1x4x319 & 524.0390 & 1.206 & DD & -2386 ± 72 & -26 ± 41 &  &  \\
 &  & 2022-02-18 & 2459629.9653 & ESPaDOnS & 1x4x319 & 527.5670 & 1.205 & DD & -2841 ± 70 & -50 ± 38 &  &  \\
 &  & 2022-02-21 & 2459632.8058 & ESPaDOnS & 1x4x319 & 523.3311 & 1.206 & DD & -947 ± 44 & 41 ± 35 &  &  \\
 &  & 2022-02-22 & 2459633.9703 & ESPaDOnS & 1x4x319 & 528.3550 & 1.205 & DD & -2115 ± 70 & 40 ± 45 &  &  \\
HD 63843 & 35884762 & 2024-01-18 & 2460329.0059 & ESPaDOnS & 1x4x280 & 526.3450 & 1.189 & DD & 2191 ± 15 & 5 ± 10 & >6853 ± 47 & Unknown \\
HD 68351 & 97312819 & 2012-12-23 & 2456286.1349 & ESPaDOnS & 1x4x230 & 517.1605 & 1.205 & DD & 43 ± 8 & -11 ± 9 & >253 ± 39 & Unknown \\
 &  & 2012-12-25 & 2456288.1286 & ESPaDOnS & 1x4x230 & 517.5697 & 1.205 & DD & 78 ± 12 & 7 ± 12 &  &  \\
HD 73857 & 366632312 & 2018-04-18 & 2458227.3810 & Narval & 1x4x900 & 541.1414 & 1.191 & DD & -3.1 ± 8.5 & 13.4 ± 8.6 & >10 ± 26 & Unknown \\
HD 81009 & 61004258 & 1999-01-14 & 2451193.6448 & MuSiCoS & 1x4x360 & 532.6345 & 1.196 & DD & 2412 ± 60 & -15 ± 34 & >7555 ± 189 & Unknown \\
HD 188774 & 171095675 & 2014-09-06 & 2456907.9513 & ESPaDOnS & 1x4x840 & 518.4316 & 1.195 & MD & 26 ± 12 & -4 ± 13 & 258 ± 63 & Dipolar \\
 &  & 2015-07-22 & 2457227.0266 & ESPaDOnS & 10x4x129 & 516.4880 & 1.196 & DD & 53 ± 9 & 0 ± 9 &  &  \\
 &  & 2015-07-23 & 2457227.9747 & ESPaDOnS & 10x4x129 & 515.8201 & 1.197 & DD & 28 ± 10 & 3 ± 10 &  &  \\
 &  & 2015-07-24 & 2457228.8548 & ESPaDOnS & 10x4x129 & 516.0640 & 1.196 & DD & -55 ± 9 & -9 ± 9 &  &  \\
 &  & 2015-07-25 & 2457229.9180 & ESPaDOnS & 20x4x129 & 519.1700 & 1.195 & MD & 55 ± 10 & -8 ± 10 &  &  \\
 &  & 2015-07-27 & 2457232.0566 & ESPaDOnS & 10x4x129 & 519.7395 & 1.195 & DD & -36 ± 11 & -6 ± 11 &  &  \\
 &  & 2015-07-28 & 2457232.9331 & ESPaDOnS & 15x4x129 & 517.2552 & 1.196 & DD & 72 ± 10 & -9 ± 10 &  &  \\
 &  & 2015-07-31 & 2457236.0300 & ESPaDOnS & 10x4x129 & 519.5260 & 1.195 & DD & 84 ± 10 & -9 ± 10 &  &  \\
 &  & 2015-08-04 & 2457240.0195 & ESPaDOnS & 4x4x129 & 521.6588 & 1.195 & MD & 14 ± 22 & -32 ± 22 &  &  \\
HD 213918 & 128379228 & 2012-11-25 & 2456257.7631 & ESPaDOnS & 3x4x900 & 518.9928 & 1.203 & DD & 857 ± 85 & 65 ± 82 & 10880 ± 252 & Quadrupolar \\
 &  & 2012-11-28 & 2456260.7471 & ESPaDOnS & 1x4x900 & 519.6083 & 1.203 & DD & -599 ± 64 & 42 ± 58 &  &  \\
 &  & 2012-11-29 & 2456261.7982 & ESPaDOnS & 2x4x900 & 519.7885 & 1.203 & DD & 1605 ± 99 & 26 ± 98 &  &  \\
 &  & 2012-12-03 & 2456265.7077 & ESPaDOnS & 1x4x900 & 520.7785 & 1.203 & DD & 3353 ± 78 & -42 ± 70 &  &  \\
 &  & 2012-12-06 & 2456268.7126 & ESPaDOnS & 1x4x900 & 520.1573 & 1.203 & DD & 2905 ± 74 & 84 ± 67 &  &  \\
$\rho$\,Pup & 154360594 & 2014-02-09 & 2456698.8667 & ESPaDOnS & 2x4x30 & 548.9392 & 1.187 & ND & -0.1 ± 1 & -0.3 ± 1 & 0.9 ± 2.8 & UW fossil \\
 &  & 2015-10-30 & 2457327.1443 & ESPaDOnS & 8x4x25 & 541.0814 & 1.188 & DD & -0.3 ± 0.9 & 0.4 ± 0.9 &  & \\
\hline
\end{tabular}}
\label{tab:specpol}
\end{table}
}

\newpage

{\onecolumn
\begin{table}[ht!]
\centering
\caption{Summary of spectropolarimetric observations for the sample. }
\resizebox{\textwidth}{!}{%
\begin{tabular}{lcccccccc}
\hline
\multicolumn{1}{c}{1}\Tstrut & 2 & 3 & 4 & 5 & 6 & 7 & 8 & 9 \\
\multicolumn{1}{c}{ID} & Spectral Type & m$_v$ & Mass & Radius & T$_{\rm eff}$ & Parallax & L & Age \\
 & \multicolumn{1}{l}{} & \multicolumn{1}{l}{} & ($M_\odot$) & ($R_\odot$) & (K) & (mas) & ($L_\odot$) & (Myr) \\ \hline
$\beta$\,Cas$^1$ & F2III & 2.27 & 2.09 ± 0.02 & 3.56 ± 0.13 & 7080 ± 20 & 59.58 ± 0.38$^\dagger$ & 30.3 ± 0.4$^\dagger$ & 1127 ± 49$^\dagger$ \\
HD 8441$^2$ & A4VpSrSi & 6.68 & 2.76 ± 0.18 & 3.03 ± 0.25 & 9200 ± 100 & 3.39 ± 0.07$^\dagger$ & 161.5 ± 6.7$^\dagger$ & 374 ± 0$^\dagger$ \\
HD 10783$^3$ & A2SiCrSr & 6.43 & 2.72 ± 0.13 & 2.57 ± 0.25$^\dagger$ & 10351 ± 500 & 5.91 ± 0.09$^\dagger$ & 66.9 ± 2$^\dagger$ & 294 ± 48$^\dagger$ \\
HD 36955 & kA1mA3V & 9.58 & 2.3 ± 0.1$^\dagger$ & 1.85 ± 0.05$^\dagger$ & 9605 ± 80$^\dagger$ & 2.23 ± 0.04$^\dagger$ & 25.7 ± 1$^\dagger$ & 233 ± 6$^\dagger$ \\
HD 41641$^4$ & A5III & 7.86 & 2.3 ± 0.6 & 4.5 ± 2.2 & 7200 ± 80 & 5.27 ± 0.05$^\dagger$ & 22.5 ± 0.5$^\dagger$ & 1433 ± 38$^\dagger$ \\
HD 49198 & A0III-IVCrSi & 9.31 & 2.56 ± 0.03$^\dagger$ & 2.29 ± 0.14 & 9839 ± 220$^\dagger$ & 2.10 ± 0.04$^\dagger$ & 37.4 ± 1.4$^\dagger$ & 158 ± 9$^\dagger$ \\
HD 63843 & A2IVSrSi & 10.25 & 3.34 ± 0.05$^\dagger$ & 5.12 ± 0.3$^\dagger$ & 8000 ± 205$^\dagger$ & 0.85 ± 0.02$^\dagger$ & 94.9 ± 5.2$^\dagger$ & 565 ± 18$^\dagger$ \\
HD 68351$^5$ & A0VpSiCr & 5.61 & 3.33 ± 0.14 & 5.15 ± 0.25$^\dagger$ & 9925 ± 60 & 4.67 ± 0.22$^\dagger$ & 227.5 ± 21.8$^\dagger$ & 298 ± 23$^\dagger$ \\
HD 73857$^6$ & A9III & 7.18 & 1.6 ± 0.1 & 2.9 ± 0.435 & 6909 ± 220 & 4.51 ± 0.03$^\dagger$ & 15.96 ± 0.7 & 751 ± 16$^\dagger$ \\
HD 81009$^7$ & ApEuCrSr & 6.53 & 2.2 ± 0.2 & 2.6 ± 0.5 & 8500 ± 250 & 6.92 ± 0.61$^\dagger$ & 44.4 ± 7.8$^\dagger$ & 613 ± 54$^\dagger$ \\
HD 188774$^8$ & A9IV & 8.81 & 2.61 ± 0.10 & 3.57 ± 0.5355 & 7600 ± 30 & 2.40 ± 0.02$^\dagger$ & 45.1 ± 0.6$^\dagger$ & 778 ± 42$^\dagger$ \\
HD 213918 & B6IVpSiSrFe & 8.68 & 3.07 ± 0.03$^\dagger$ & 2.57 ± 0.1$^\dagger$ & 14297 ± 255$^\dagger$ & 2.03 ± 0.03$^\dagger$ & 243.6 ± 8.0$^\dagger$ & 48 ± 7$^\dagger$ \\
$\rho$\,Pup$^9$ & F5IIkF2IImF5II & 2.81 & 1.85 ± 0.20 & 3.52 ± 0.07 & 6675 ± 175 & 51.40 ± 0.24$^\dagger$ & 24.8 ± 0.2$^\dagger$ & 1177 ± 44$^\dagger$ \\ \hline
\end{tabular}}
\vskip\baselineskip
\begin{tabular}{lccccccccccc}
\hline
\Tstrut & 10 & 11 & 12 & 13 & \revision{14} & 15 & 16 & 17 & 18 & 19 & 20 \\
\multicolumn{1}{c}{ID} & $v\sin i$ & $i$ & Period & $v_{\rm rot}$ & \revision{$\Omega/\Omega_{\rm crit}$} & u & B$_{\rm pol}$ & B$_{\rm c,s,Z}$ & B$_{\rm c,s,S}$ & B$_{\rm pol,ZAMS}$ & Mag. field \\
 & (km s$^{-1}$) & ${}^\circ$ & (d) & (km s$^{-1}$) & \revision{(\%)} & \multicolumn{1}{l}{} & (G) & (G) & (G) & (G) & \multicolumn{1}{l}{} \\ \hline
$\beta$\,Cas$^1$ & 70 & 20 & 0.868 & 205 & \revision{62} & 0.5902 & 17 ± 9 & 1 & 92 & 84 & D \\
HD 8441$^2$ & 2 & 65 & 69.430 & 2 & \revision{1} & 0.5151 & $>$180 ± 104 & 2 & 107 & 1313 & U \\
HD 10783$^3$ & 20 & 40 & 4.134 & 31 & \revision{7} & 0.4797 & $>$1906 ± 44 & 3 & 54 & 3679 & U \\
HD 36955 & 33 & 51 & 2.284 & 42 & \revision{8} & 0.5046 & $>$1849 ± 191 & 2 & 112 & 1953 & U \\
HD 41641$^4$ & 30 & 76 & 5.615 & 31 & \revision{13} & 0.5785 & 748 ± 28 & 1 & 18 & 2414 & F \\
HD 49198 & 15 & 53 & 6.224 & 19 & \revision{4} & 0.4965 & 11740 ± 2700 & 1 & 46 & 11991 & F \\
HD 63843 & 7 & 3 & 1.670 & 134 & \revision{44} & 0.5751 & $>$6853 ± 47 & 1 & 20 & 41772 & U \\
HD 68351$^5$ & 54 & 24 & 6.494 & 40 & \revision{11} & 0.4876 & $>$253 ± 39 & 1 & 31 & 1728 & U \\
HD 73857$^6$ & 9 & 3 & 0.716 & 172 & \revision{63} & 0.5902 & $>$10 ± 26 & 2 & 216 & 35 & U \\
HD 81009$^7$ & 3 & 139 & 33.972 & 5 & \revision{1} & 0.5630 & $>$7555 ± 189 & 1 & 12 & 25228 & U \\
HD 188774$^8$ & 52 & 50 & 2.907 & 68 & \revision{17} & 0.5916 & 258 ± 63 & 1 & 49 & 1173 & F \\
HD 213918 & 52 & 35 & 1.431 & 91 & \revision{19} & 0.3891 & 10880 ± 252 & 2 & 117 & 16708 & F \\
$\rho$\,Pup$^9$ & 8 & 16 & 6.557 & 29 & \revision{9} & 0.6030 & 0.9 ± 2.8 & 2 & 151 & 5 & F \\ \hline
\end{tabular}
\vskip\baselineskip
\begin{tablenotes}
      \item \textbf{Notes.} Aside from those determined through our own analysis (i.e. $v\sin i$, $u$, B$_{\rm pol}$), parameters are retrieved from [1] \citet{zwintz2020}, [2] \citet{north1998}, [3] \citet{netopil2017}, [4] \citet{escorza2016}, [5] \citet{wraight2012}, [6] \citet{casagrande2011}, [7] \citet{wade2000}, [8] \citet{lampens2013}, [9] \citet{neiner2017}, or alternatively [$\dagger$] inferred from \emph{Gaia} \citep{gaia2023}. The field characterisations shown in the last column correspond to dynamo (D), fossil (F), or as-of-yet unknown (U).
\end{tablenotes}
\label{tab:params}
\end{table}
}


\newpage

{\onecolumn
\section{Figures} \label{app:figs}
\begin{figure}[!ht]
    \centering
    \begin{subfigure}{0.9\textwidth}
         \centering
         \includegraphics[width=\textwidth]{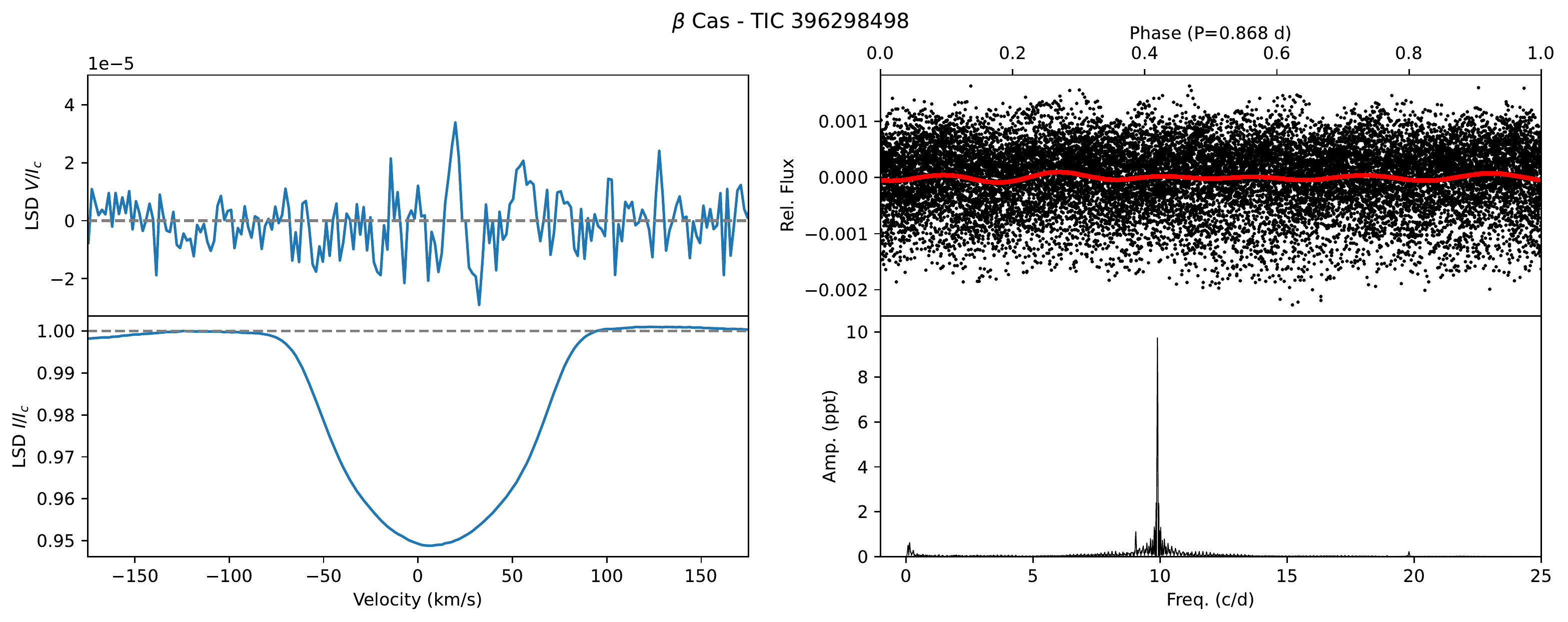}
         \label{fig:betacas_lsd}
     \end{subfigure}
     \vskip\baselineskip
     \begin{subfigure}{0.9\textwidth}
         \centering
         \includegraphics[width=\textwidth]{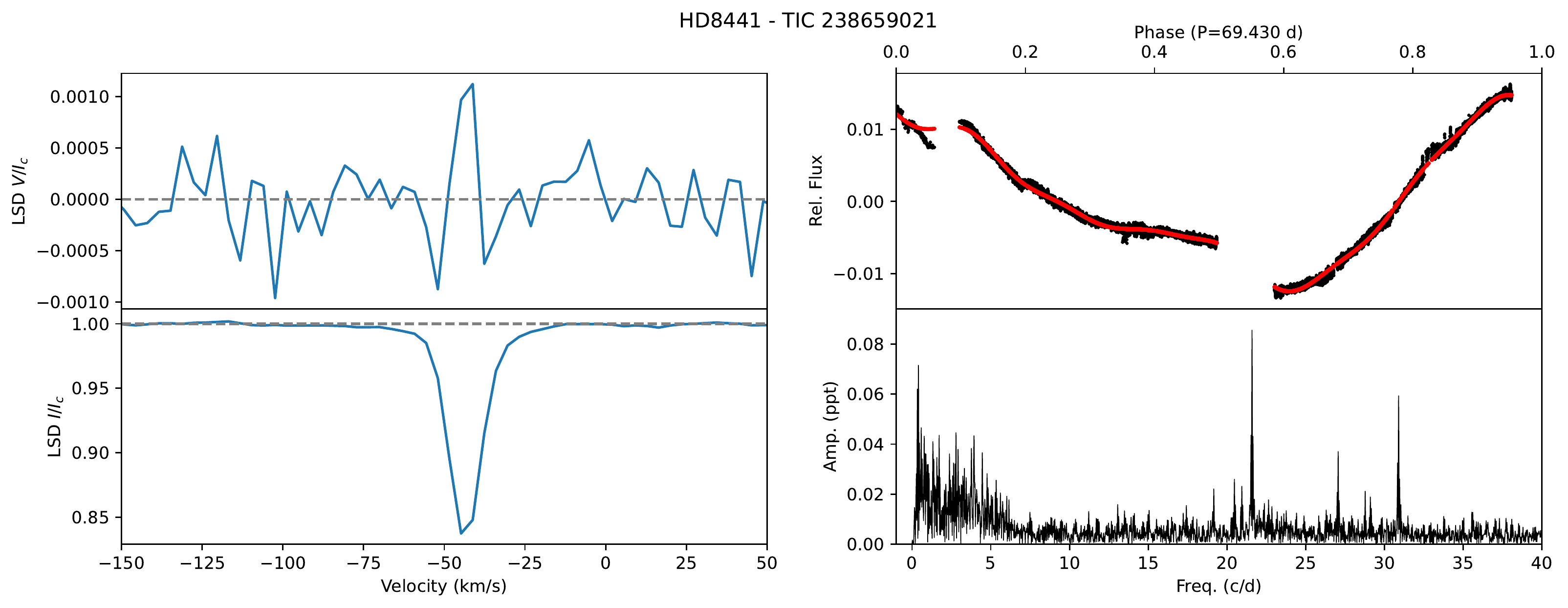}
         \label{fig:hd8441_lsd}
     \end{subfigure}
     \vskip\baselineskip
     \begin{subfigure}{0.9\textwidth}
         \centering
         \includegraphics[width=\textwidth]{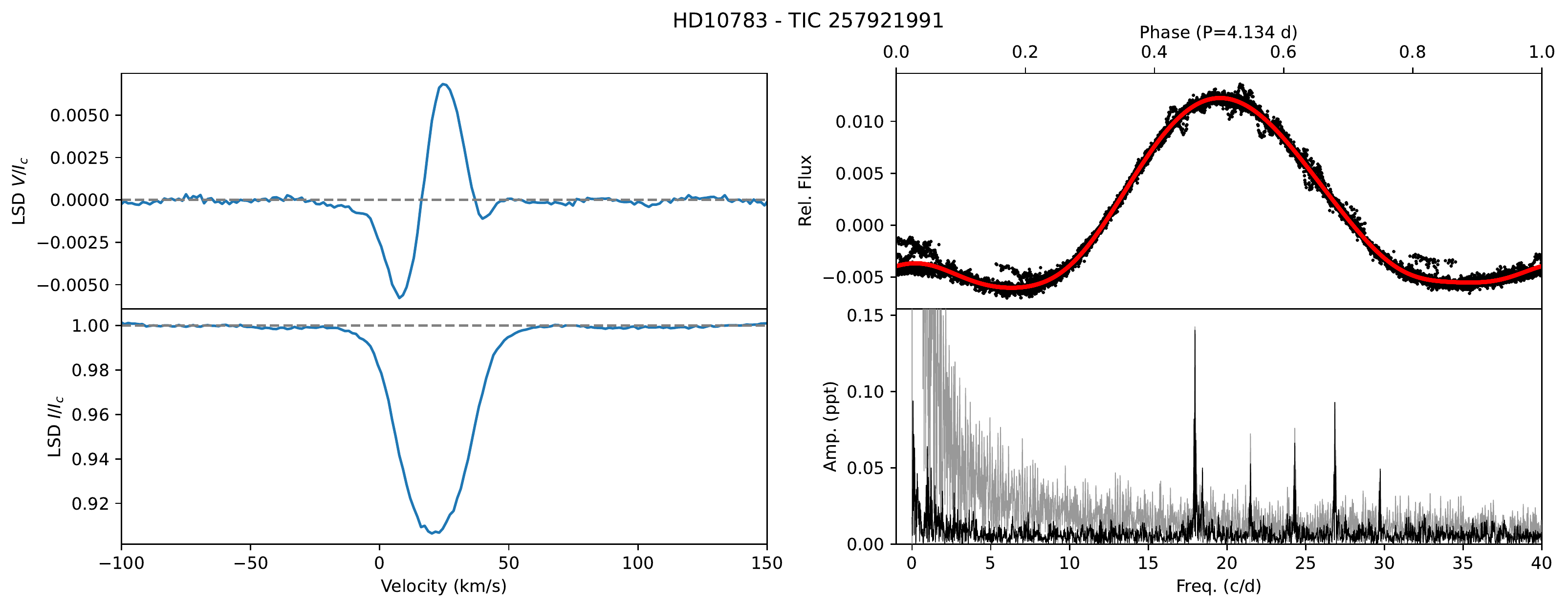}
         \label{fig:hd10783_lsd}
     \end{subfigure}
    \caption{LSD Stokes V (upper-left panel), Stokes I (lower-left panel), and Lomb-Scargle periodograms for two different TESS sectors (right panel) for the various stars in our sample, arranged by stellar identifier. For stars with a large amount of low-frequency amplitude the lightcurves were pre-whitened, removing low-frequency high-amplitude peaks, resulting in the before (grey) and after (black) periodograms. In each case, the x-axis extends to cover the highest frequency signals found in the data. \revision{The red curve is a 5-term fit to the rotation frequency}.}
    \label{fig:stokes}
\end{figure}

\begin{figure}[!ht]
    \centering
     \begin{subfigure}{0.9\textwidth}
         \centering
         \includegraphics[width=\textwidth]{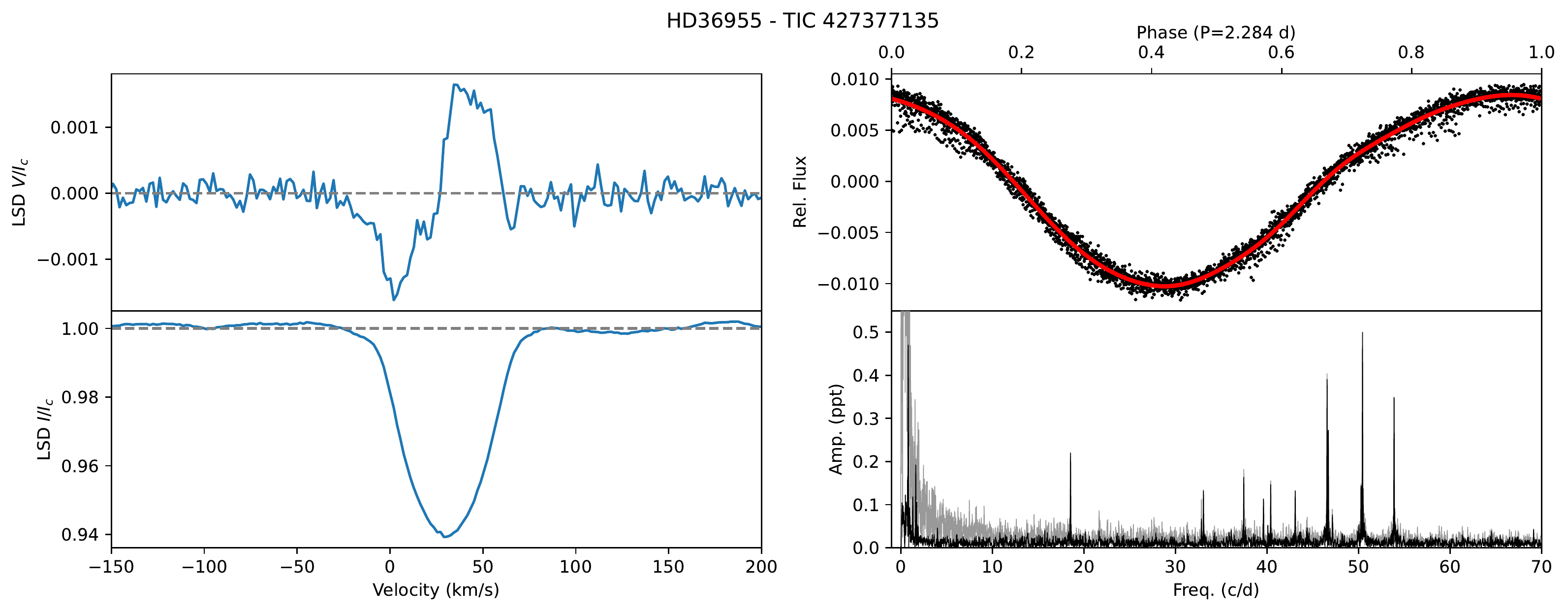}
         \label{fig:hd36955_lsd}
     \end{subfigure}
     \vskip\baselineskip
     \begin{subfigure}{0.9\textwidth}
         \centering
         \includegraphics[width=\textwidth]{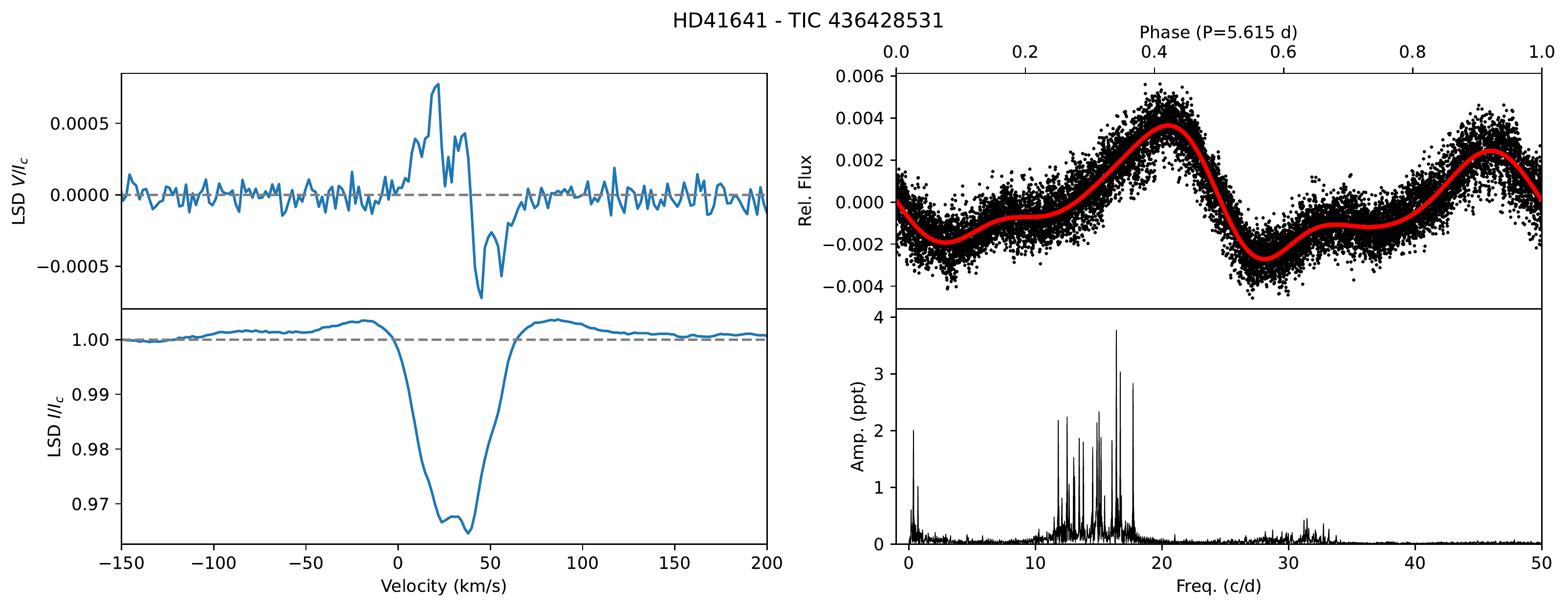}
         \label{fig:hd41641_lsd}
     \end{subfigure}
     \vskip\baselineskip
     \begin{subfigure}{0.9\textwidth}
         \centering
         \includegraphics[width=\textwidth]{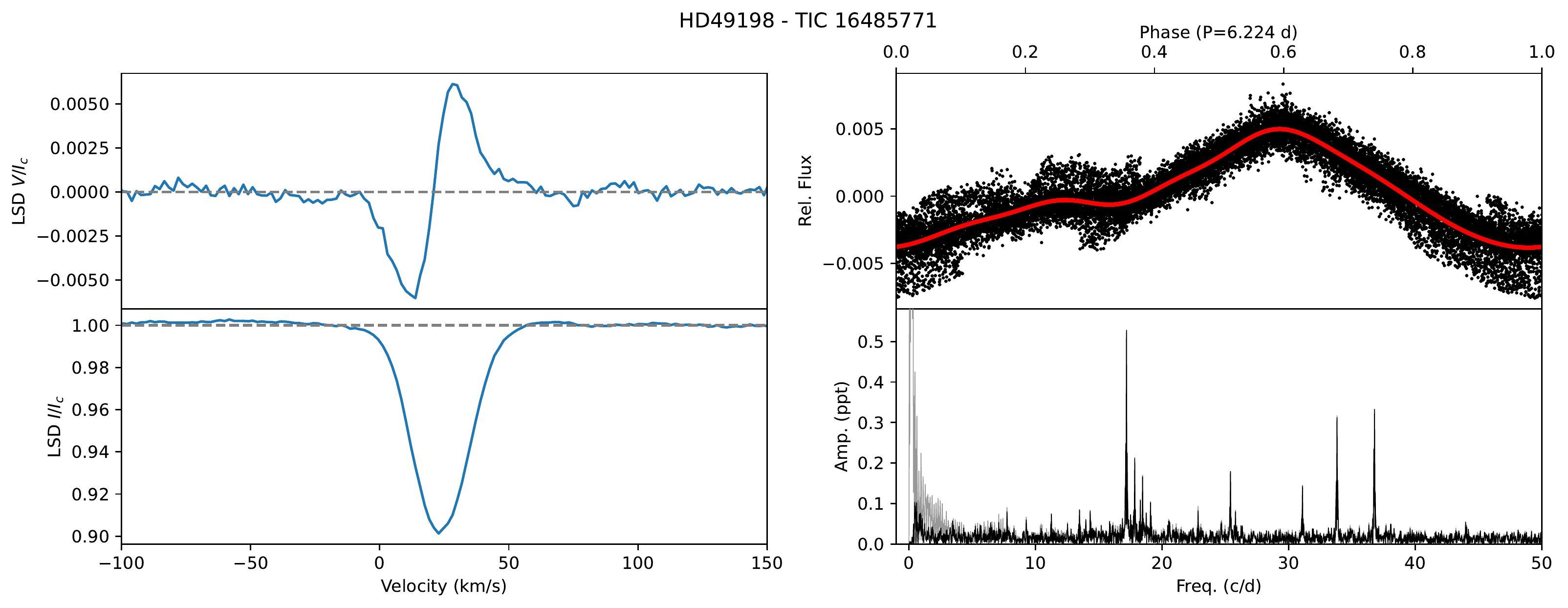}
         \label{fig:hd49198_lsd}
     \end{subfigure}
    \caption{LSD Stokes V (upper-left panel), Stokes I (lower-left panel), and Lomb-Scargle periodograms for two different TESS sectors (right panel) for the various stars in our sample, arranged by stellar identifier. For stars with a large amount of low-frequency amplitude the lightcurves were pre-whitened, removing low-frequency high-amplitude peaks, resulting in the before (grey) and after (black) periodograms. In each case, the x-axis extends to cover the highest frequency signals found in the data.  \revision{The red curve is a 5-term fit to the rotation frequency} (cont.).}
    \label{fig:stokes2}
\end{figure}

\begin{figure}[!ht]
    \centering
     \begin{subfigure}{0.9\textwidth}
         \centering
         \includegraphics[width=\textwidth]{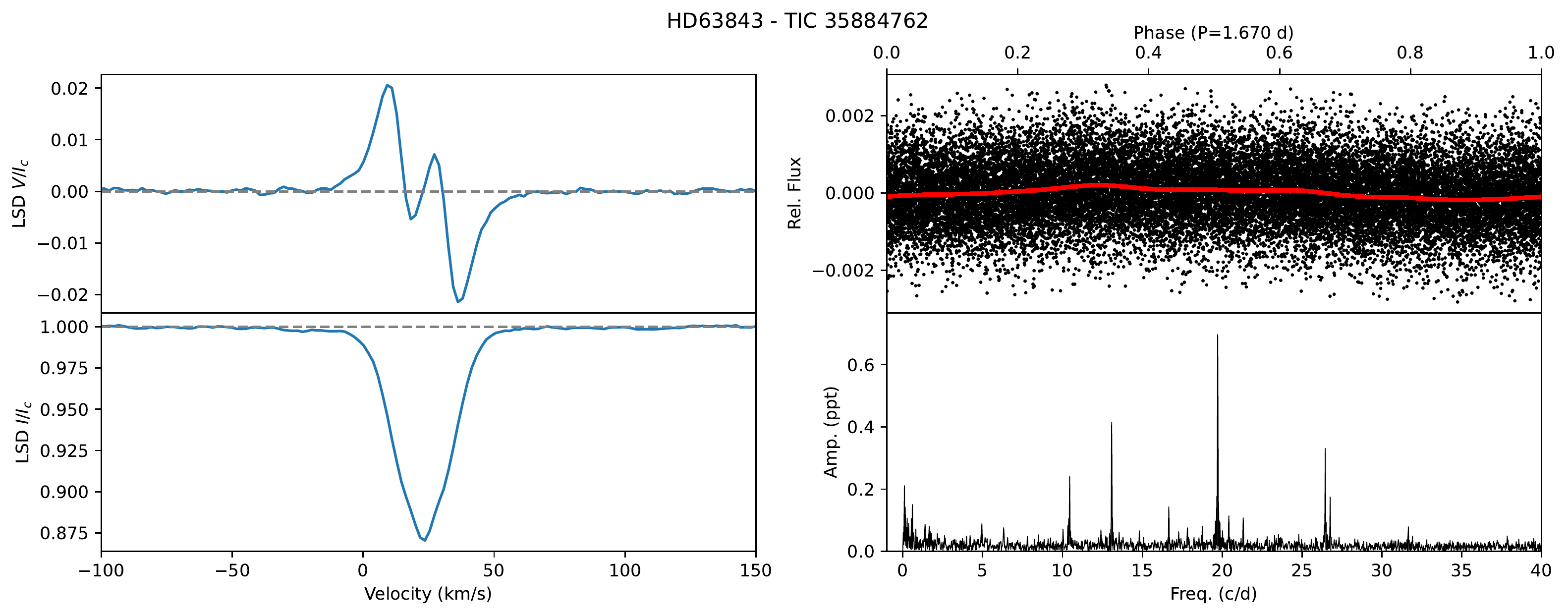}
         \label{fig:hd63843_lsd}
     \end{subfigure}
     \vskip\baselineskip
     \begin{subfigure}{0.9\textwidth}
         \centering
         \includegraphics[width=\textwidth]{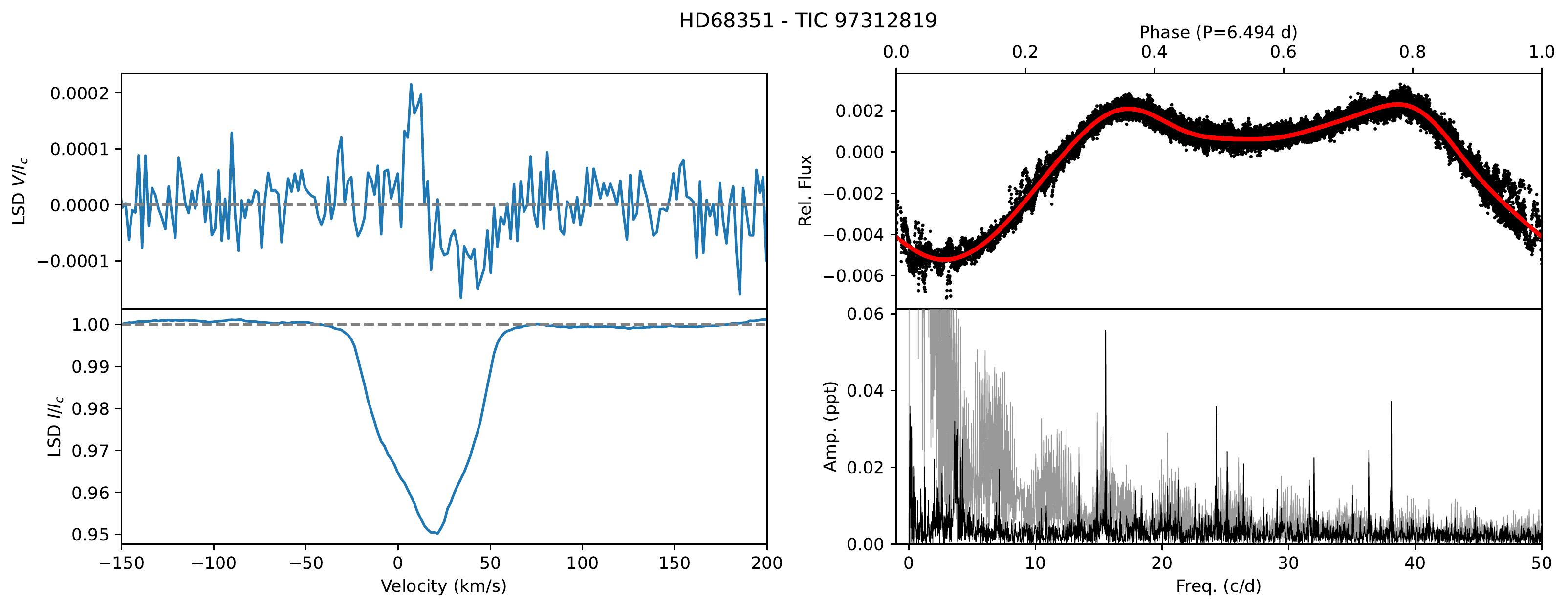}
         \label{fig:hd68351_lsd}
     \end{subfigure}
     \vskip\baselineskip
     \begin{subfigure}{0.9\textwidth}
         \centering
         \includegraphics[width=\textwidth]{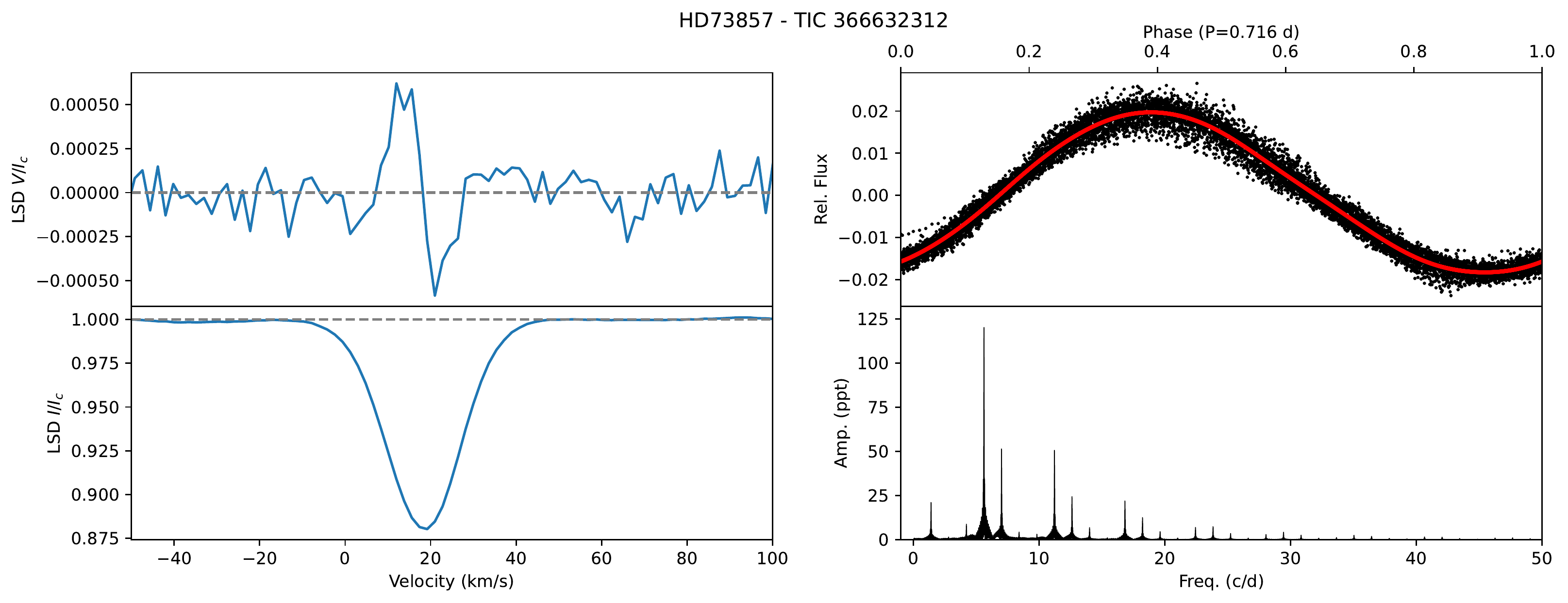}
         \label{fig:hd73857_lsd}
     \end{subfigure}
    \caption{LSD Stokes V (upper-left panel), Stokes I (lower-left panel), and Lomb-Scargle periodograms for two different TESS sectors (right panel) for the various stars in our sample, arranged by stellar identifier. For stars with a large amount of low-frequency amplitude the lightcurves were pre-whitened, removing low-frequency high-amplitude peaks, resulting in the before (grey) and after (black) periodograms. In each case, the x-axis extends to cover the highest frequency signals found in the data. \revision{The red curve is a 5-term fit to the rotation frequency} (cont.).}
    \label{fig:stokes3}
\end{figure}

\begin{figure}[ht!]
    \centering
     \begin{subfigure}{0.9\textwidth}
         \centering
         \includegraphics[width=\textwidth]{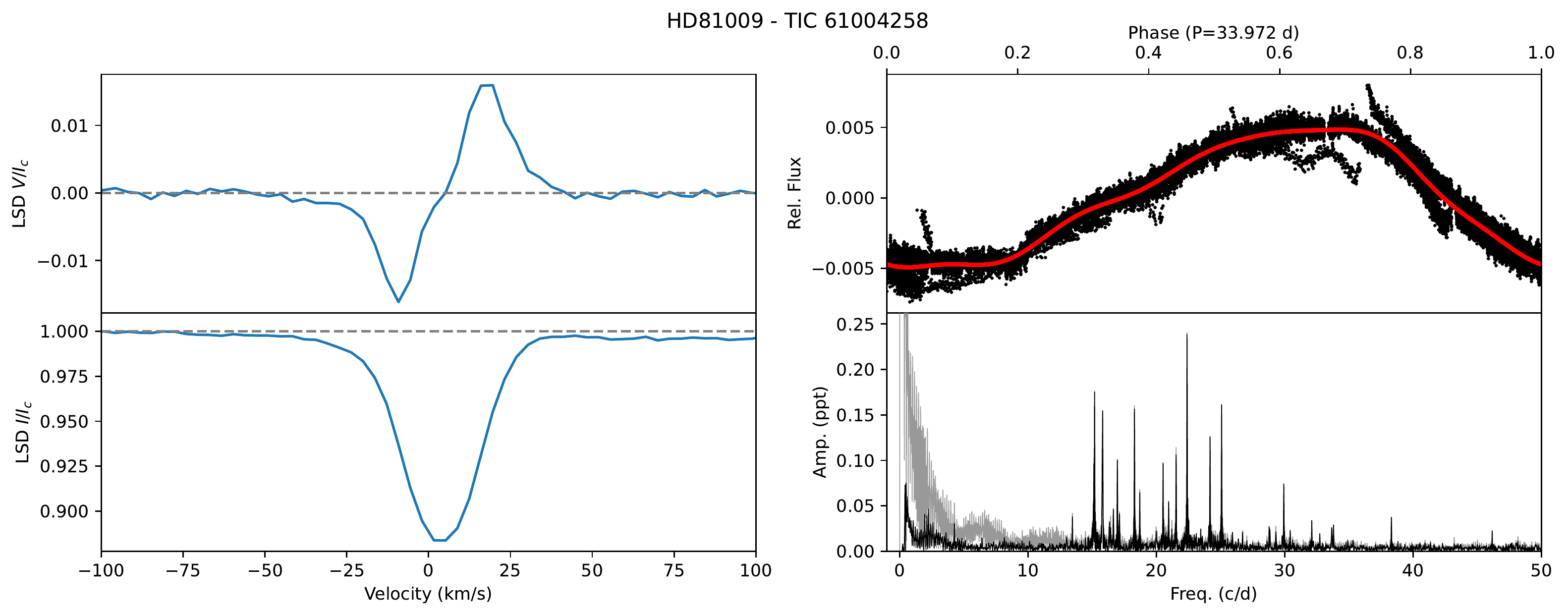}
         \label{fig:hd81009_lsd}
     \end{subfigure}
     \vskip\baselineskip
     \begin{subfigure}{0.9\textwidth}
         \centering
         \includegraphics[width=\textwidth]{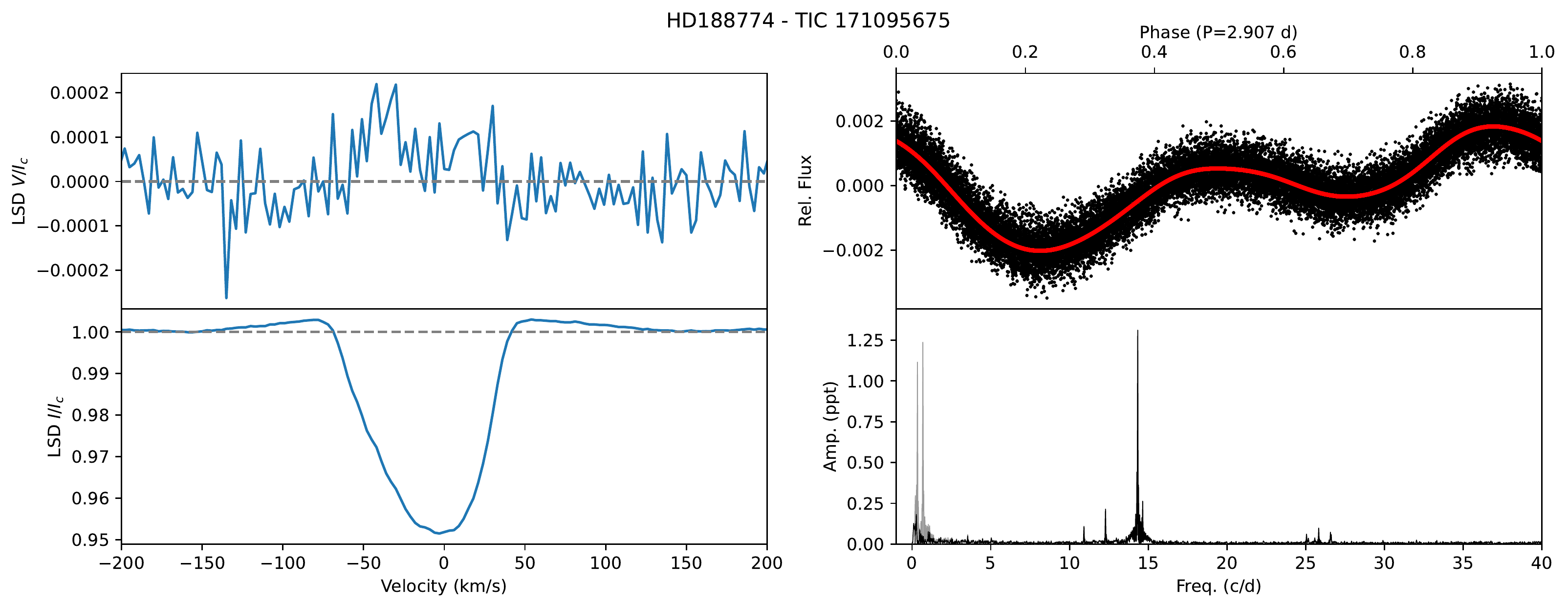}
         \label{fig:hd188774_lsd}
     \end{subfigure}
     \vskip\baselineskip
     \begin{subfigure}{0.9\textwidth}
         \centering
         \includegraphics[width=\textwidth]{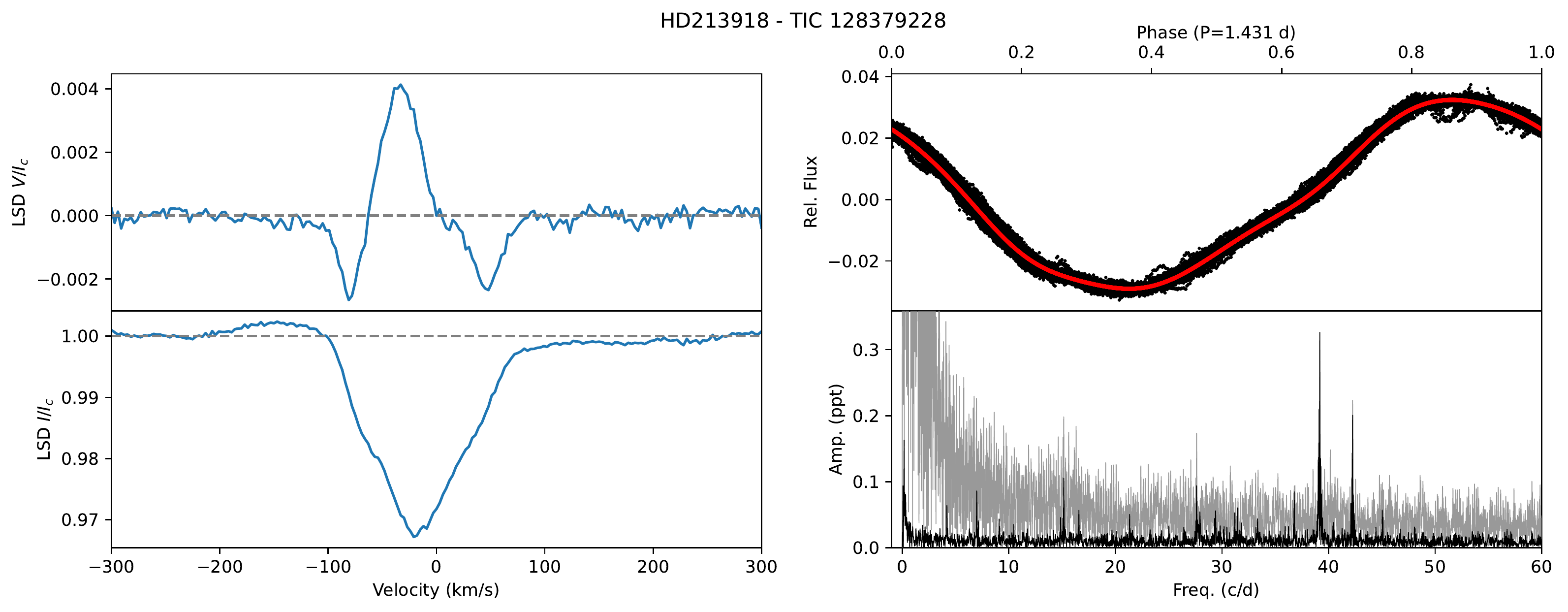}
         \label{fig:hd213918_lsd}
     \end{subfigure}
    \caption{LSD Stokes V (upper-left panel), Stokes I (lower-left panel), and Lomb-Scargle periodograms for two different TESS sectors (right panel) for the various stars in our sample, arranged by stellar identifier. For stars with a large amount of low-frequency amplitude the lightcurves were pre-whitened, removing low-frequency high-amplitude peaks, resulting in the before (grey) and after (black) periodograms. In each case, the x-axis extends to cover the highest frequency signals found in the data. \revision{The red curve is a 5-term fit to the rotation frequency} (cont.).}
    \label{fig:stokes4}
\end{figure}

\begin{figure}[ht!]
    \centering
    \begin{subfigure}{0.9\textwidth}
         \centering
         \includegraphics[width=\textwidth]{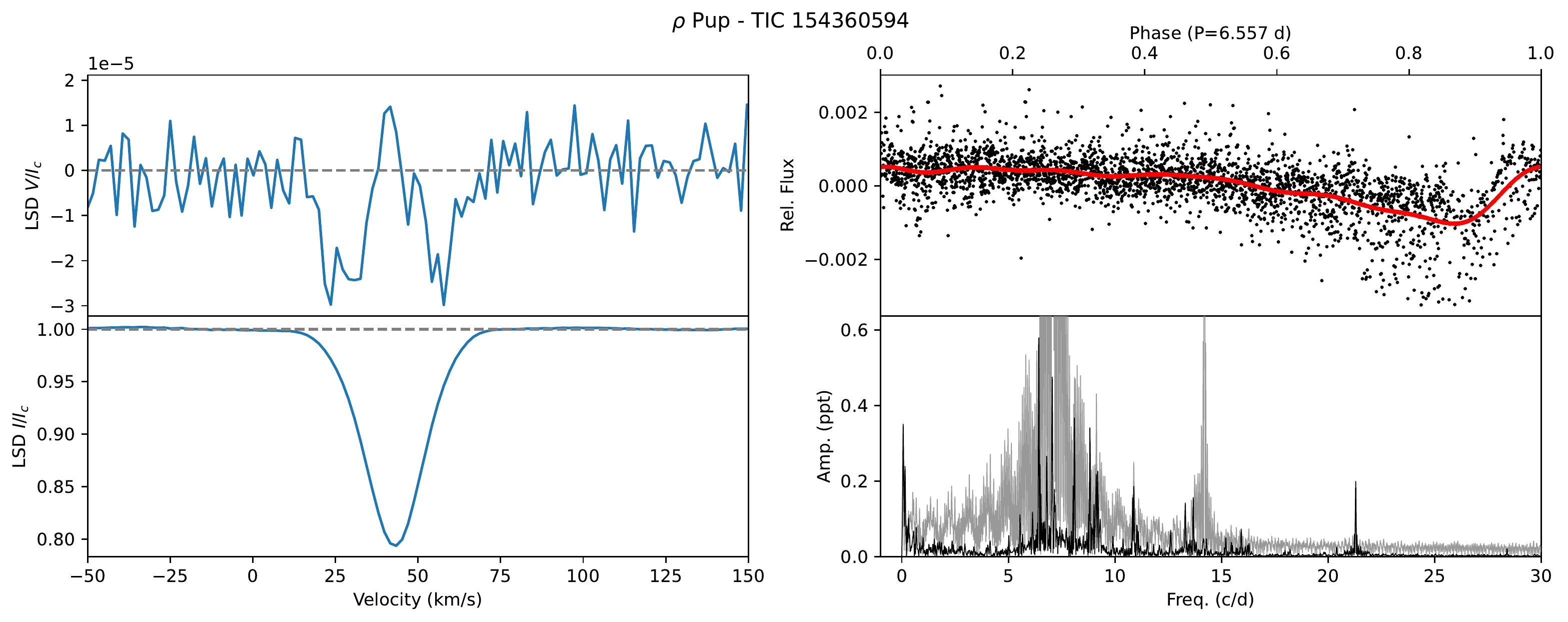}
         \label{fig:rhopup_lsd}
     \end{subfigure}
    \caption{LSD Stokes V (upper-left panel), Stokes I (lower-left panel), and Lomb-Scargle periodograms (right panel) for the various stars in our sample, arranged by stellar identifier. For stars with a large amount of low-frequency amplitude the lightcurves were pre-whitened, removing low-frequency high-amplitude peaks, resulting in the before (grey) and after (black) periodograms. In each case, the x-axis extends to cover the highest frequency signals found in the data. \revision{The red curve is a 5-term fit to the rotation frequency} (cont.).}
    \label{fig:stokes5}
\end{figure}

\begin{figure}
    \centering
    \includegraphics[width=0.9\linewidth]{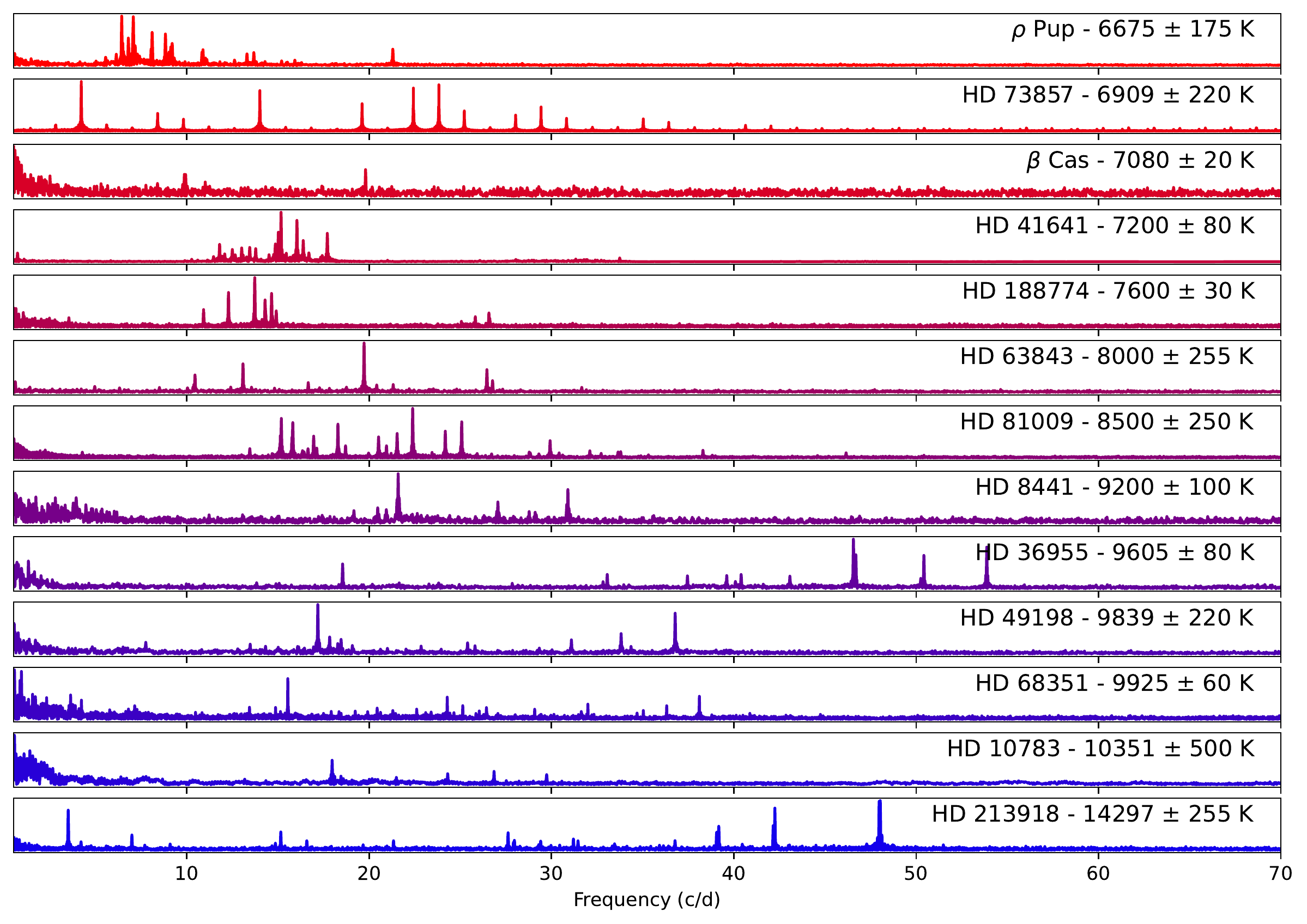}
    \caption{Lomb-Scargle periodograms for the stars in our sample, similar to Fig.~\ref{fig:powerspectrum}, except in this case presented in descending order with respect to the effective temperature $T_{\text{eff}}$.}
    \label{fig:ps_teff}
\end{figure}
}

\end{appendix}

\end{document}